\documentclass[preprint]{aastex}
\newcommand{\kms}{\ensuremath{\mathrm{km\ s}^{-1}}}

\newcommand\Halpha{H$\alpha$}
\newcommand\Hbeta{H$\beta$}
\newcommand\Shb{S(H$\beta$)}

\newcommand\funits{$\rm erg~cm^{-2}~s^{-1}$}
\newcommand\cmq{$\rm cm^{-3}$}
\newcommand\cmsq{$\rm cm^{-2}$}

\newcommand\ori{$\theta ^{1}$Ori~C}
\newcommand\oriA{$\theta ^{2}$Ori~A}

\newcommand\Te{T$\rm _{e}$}
\newcommand\Ne{N$\rm _{e}$}

\newcounter{ionstage}

\newcommand\chb{c$\rm _{H\beta}$}
\newcommand\ps{s$\rm ^{-1}$}
\newcommand\arcsecsq{arcsec$\rm ^{-2}$}
\slugcomment{Accepted for Publication in the Astronomical Journal}
\shorttitle{Spectrophotometry of the Orion Nebula Region}
\shortauthors{O'Dell \& Goss}

\begin{document}

\title{Spectrophotometry of the Huygens Region of the Orion Nebula, the Extended Orion Nebula, and M~43; Scattered Light Systematically Distorts Conditions Derived from Emission-Lines
\footnote{
Based in part on observations obtained at the Cerro Tololo Inter-American Observatory, which is operated by the Association of Universities for Research in Astronomy, Inc., under a Cooperative Agreement with the National Science Foundation.}}


\author{C. R. O'Dell}
\affil{Department of Physics and Astronomy, Vanderbilt University, Box 1807-B, Nashville, TN 37235}
\email{cr.odell@vanderbilt.edu}
\and
\author{Jessica A. Harris}
\affil{Department of Physics, Fisk University, 1000 17th Avenue, Nashville, TN 37208}
\email{jessica.a.harris@vanderbilt.edu}

\begin{abstract}

We report on medium resolution spectrophotometry of the Orion Nebula region, including for the first time the Extended Orion Nebula and the nearby M~43. The 49 long slit observations were divided into 99 smaller samples, which have allowed determinations of the amount of extinction, extinction corrected \Hbeta\ surface brightness, electron temperatures (from [S~II], [N~II], and [O~III]), and electron densities (from [S~II] and [Cl~III]) throughout much of this complex region.

We verify an earlier conclusion from a radio/optical study that beyond about 5\arcmin\ from \ori\  local emission begins to be contaminated by scattering of light from the much brighter central Huygens Region of M~42 and this scattered light component becomes dominant at large distances. This contamination means that the derived properties for the outer regions are not accurate. From comparison of the light from the dominant star in M~43 with the continuum of that nebula (which is almost entirely scattered star light) it is determined that scattered light is enhanced in the blue, which can lead to observed Balmer line ratios that are theoretically impossible and erroneous electron temperatures. 

This blue scattering of emission-lines is important even in the Huygens Region because it means that at anything except very high spectroscopic resolution the observed lines are a blend of the original and scattered light, with shorter wavelength lines being artificially enhanced.  This can lead to over-estimates of the electron temperatures derived from the nebular and auroral line ratios of forbidden lines.  This phenomenon is probably applicable to many other H~II regions.

We have been able to use extinction-insensitive line ratios, the extinction corrected surface brightness in \Hbeta, and the equivalent width of the continuum to create for the first time a three dimensional model of the entire M~42, Extended Orion Nebula, and M~43 region.  This is an irregular concave blister of ionized gas bounded on the outside by apparent walls where the ionization front has curved almost to the direction of the observer. M~43 is seen to be shielded from illumination by \ori\ by the northeast portion of the wall bounding M~42.

\end{abstract}

\keywords{Galactic Nebulae:individual(Orion Nebula, NGC1976, NGC 1982, M 42, M 43)}

\section{Introduction\label{sect:intro}}

The Orion Nebula (NGC~1976, M~42) is commonly thought of as a small HII region, the bright Huygens Region \citep{gin} being only about  0.7 pc (5\arcmin\ at a distance of 440 pc \citep{oh08}, the distance adopted in this study). However, when one includes the much fainter Extended Orion Nebula (EON, a term introduced by \citep{gud}), it becomes comparable to other Galactic and extragalactic HII regions since the angular size of  about 33.3\arcmin\ x 23.8\arcmin\ corresponds to about 4.3 pc x 3.1 pc.  The Huygens Region occupies the northeast corner of the EON and has been frequently imaged in narrow band emission-line filters, event at Hubble Space Telescope (HST) resolution \citep{ow96} whereas the EON has largely been the provence of skilled amateur astronomers. Much of the EON has been imaged by the HST with filters typically passing several emission- lines \citep{wjh}, while an excellent example of a full-field ground telescope image is that of Robert Gendler, M.D. (http://www.robgendlerastropics.com). The wide field of view images often include the nearby low ionization H~II Region M~43 (NGC~1982) which lies to the northeast of M~42 and this object has not been well studied but is included in this investigation.

Because the EON is dominated by regions about two orders of magnitude lower surface brightness than the Huygens region, it has not been the subject of detailed spectroscopic study.  Whereas there have been many high spectral resolution spectrophotometric studies of the Huygens region, e.g. \citet{b00},\citet{est04} and one Fabry-Perot study of the 3727 \AA\ [O~II] doublet of the entire nebula \citep{ca72}, only a few lower resolution studies have been made beyond a few arcminutes from \ori\  \citep{pei67,peicos69,pp77,jps73,b91,og09} and none of these cover the full EON.  In the study described in this paper we correct that lack of data of this important region. The smaller and fainter M~43 has been the subject of many fewer investigations \citep{peicos69,thu78,kha80,han87,sm87,rod99} but it's dominance by one star, NU Ori, makes it potentially an easier object to analyze.

We adopt the blister model for the Huygens Region. This model posits that the emission-lines arise from a relatively layer of ionized gas lying on the side of the Orion Molecular Cloud that faces the observer \citep{z73,bal74}.  A key distinguishing feature of this model is the difference of velocities of different species (summarized in \citet{od09a}), with the low ionization ions that give rise to [S~II] emission originating from near the main-ionization-front on the observer's side of the dense photon dominated region (PDR) having little blue-shift with respect to the parent molecular cloud. Emission in the H$^{+}$+He$^{o}$ zone, such as [N~II] \citep{od98}, is slightly blue-shifted, while emission in the H$^{+}$+He$^{+}$ zone, such as [O~III] is the most blue-shifted as the hot gas expands away from the main-ionization-front. The spectral type O7Vp \ori\ is the dominant source of ionizing photons and lies about 0.2 pc \citep{od09a} on the observer's side of the main-ionization-front (this would correspond to a distance in the plane of the sky of 1.56\arcmin).  In the foreground is a multi-component Veil of nearly neutral material \citep{vdw89} that lies about one to three parsecs in the foreground \citep{abel,od09a}. This Veil produces most of the extinction in this nearby, high Galactic latitude (-20\arcdeg),  H~II Region \citep{od92}. It is highly variable in optical depth, this being the greatest in the highly obscured Dark Bay feature lying immediately to the east of the Trapezium stars.  

Ferland \citep{b91} pointed out that a flat blister model would vary in surface brightness in the \Hbeta\ line approximately as D$\rm ^{-2}$, where D is the distance from \ori\ in the plane of the sky, in approximate agreement with the limited data then available. This approximation holds only very near the sub-stellar point and a better approximation is D$\rm ^{-3}$ since the photons strike the plane more obliquely with distances and the flux in Lyman Continuum (LyC) photons decreases correspondingly. The fact that over greater distances the surface brightness decreases more rapidly than D$\rm ^{-2}$ indicates that a more complex model applies. Determination of the location of the ionization front from the apparent surface brightness was expanded upon by \citet{wo95}, who did a detailed general solution and derived a model of the Huygens Region that is an irregular concave surface, that drops away from the observer behind the Dark Bay feature, rises abruptly at the linear Bright Bar feature running northeast-southwest to the southeast of \ori, and has a hump in the Orion-S region lying to the southwest of \ori. A subsequent study established that the Orion-S feature arises from a cloud of dense neutral material lying within the ionized gas of the Huygens Region \citep{od09a}.

Even within the Huygens Region one sees systematic variations in the ionization structure with increasing distance from \ori. These arise from the decreasing density of the gas but primarily from modification of the stellar radiation field through increasing optical depth in neutral hydrogen and helium. These numerically most abundant elements have approximately the same absorption coefficients as the much rarer heavy elements, e.g. nitrogen, oxygen, and sulfur, so that the radiation field seen by the heavy elements is determined by hydrogen and helium.  The Balmer lines arise from where-ever hydrogen is ionized, so that one expects that the observed F([O~III])/F(\Hbeta) to decrease with distance from \ori\ while the F([N~II])/F(\Halpha) and F([S~II])/F(\Halpha) ratios will increase.

It has been known for some time that the continuum radiation from the Huygens Region \citep{oh65,b91} is much stronger than expected from an ionized gas. This arises from starlight being scattered by the dense PDR that lies beyond the ionized gas and there is even a weak detection of the stellar HeII 4686 \AA\ line in the nebular continuum \citep{man72}. As measured by the equivalent width  (EW) at 4861 \AA\ (the ratio of surface brightness in \Hbeta\ to the underlying continuum). This observed parameter joins the emission-line ratios as a useful diagnostic of the conditions within the nebula and the three dimensional (3-D) structure.

This study presents a large set of new spectroscopic observations of the Huygens region and the EON (\S\ 2), derives electron temperatures and densities for the regions we sample (\S\ 3), and present useful diagnostic observed properties (\S\ 4). We then use these new observations to determine the 3-D structure of the EON and come to important conclusions about the role of wavelength dependent internal scattered light in determining the properties and abundances in this and other H~II Regions.
 
\section{Observations}\label{sect:obs}
The intent of this program was to obtain moderate resolution spectroscopic data on a over a wider range of samples within the Huygens Region and the EON  with the goal of providing accurate line ratios for the primary diagnostic lines.  This intent was expanded to also include a smaller sampling of data in Messier 43, which is associated with the B 0.5 V star NU Ori \citep{sch71}. The Boller and Chivens spectrograph on the 1.5 m telescope at the Cerro Tololo Interamerican Observatory  (CTIO) was nearly ideal for this purpose and was made available to the authors through the SMARTS consortium.  

 \subsection{The Spectroscopic Observations}
Observations were made in three sets during the 2008 and 2009 observing seasons for the Orion Nebula region. This time was nearly ideal for observing from CTIO since this was during the summer and conditions were photometrically clear during almost all the scheduled time. In order to determine the best combination of wavelength coverage and spectral resolution, two first order gratings were employed, G58 giving a scale of 2.2 \AA\ per pixel was used on 2008 November 19--23 and G09 giving 2.7 \AA\ per pixel was used on 2008 November 25 and 2009 January 16, December 9--15. The wavelength coverage was limited by the width of the Loral 1K detector used throughout and the need to avoid contamination by signal from the second spectral order. Separation of the orders was aided by using ultra-violet cutoff filters (GG 395 in November, 2008 and January, 2009 and GG 385 in December, 2009). The slit width for the first two data sets was 1.9\arcsec\ and 2.6\arcsec\ for the December, 2009 observations. Observations of stars were made with slit widths of at least 5.4\arcsec\ in order to avoid the effects of wavelength dependent atmospheric refraction. The Full Width at Half Maximum (FWHM) was 6.7 \AA\ for the GG58 observations and 7.0 \AA\ for the GG09 observations.  One pixel along the projection of the slit onto the plane of the sky was 1.30\arcsec\ for all observations.  A slit length of 429 \arcsec\ was used for the first two data sets and of 344\arcsec\ for the December, 2009 observations. 

During the first two observing runs only the photometric standard star Feige 15 was observed nightly, but at multiple zenith distances and in December, 2009 nightly Feige 15 observations were supplemented by observations of the other standard stars Feige 25 (2009 December 9--14) and Hiltner 600 (2009 December 10, 11, and 14) in order to provide a cross-check on the earlier photometric calibrations. The derived system sensitivity functions from different stars varied only a few percent over our wavelength range.  

Observations of blank fields well away from the nebula were necessary because of the extended size of the Orion Nebula complex and the fact that some of the targeted nebular lines are also present in the night sky spectra.  Two sky regions were observed, having been selected on wide-field images to be free of nebular emission. These were designated as Sky-North (5:26:15 +00:25:26) and Sky-South (5:28:18.9 -07:08:36). All positions in this paper are for the year 2000.0.  Observations of these sky regions were alternated between nebula observations and in the case of the fainter regions observed during December, 2009 these observations were typically two-thirds the exposure time of the adjacent nebula observations.

The usual pattern of CCD observing was selected. Cosmic ray event subtraction from the detector images were facilitated by making at least two exposures at each position. In the case of the brightest regions multiple exposure times were employed in order to provide reference spectra of the brightest lines, which were saturated in the longer exposures of these regions. Twilight sky observations were frequently taken to determine the flat field corrections.

\begin{figure}
\epsscale{1.0}
\plotone{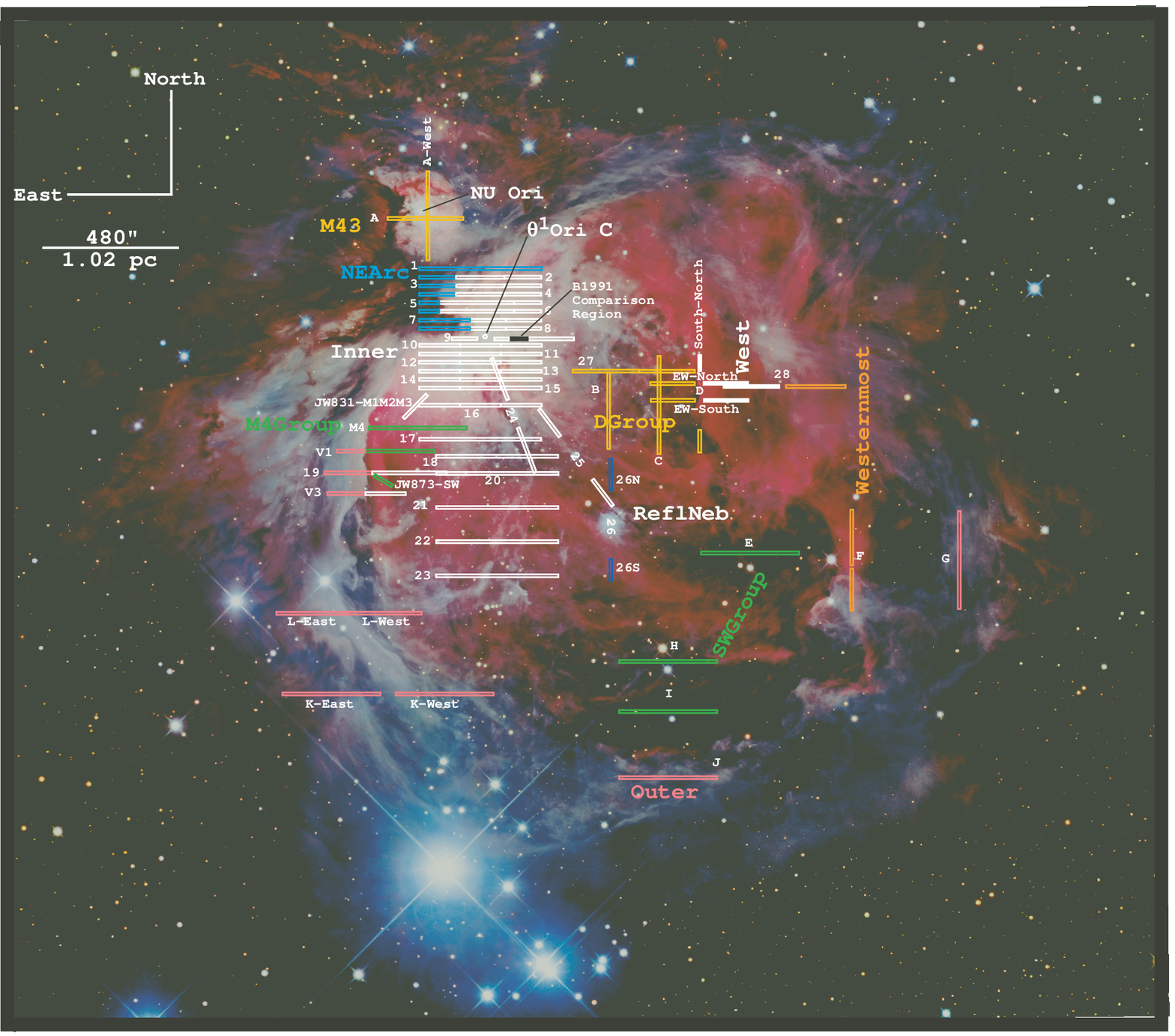}
\caption{The slit positions in this study are shown superimposed on a broad bandpass red, green, and blue ground-based image of the EON created by Robert Gendler (http://www.robgendlerastropics.com) and used with his permission. Many slit samples presented in Table 1 have been divided into subsamples named according to their orientation and position, e. g. 1-west, 1-mid, 1-east, 25SW. Groupings of samples discussed in \S\ \ref{sect:groups} are designated, with common colors indicating memberships, although the same colors are used for widely separated groups. 
\label{fig:Slits}}
\end{figure}

The slit positions were selected to give progressively wider-spaced samples in directions away from the dominant ionizing star \ori. These positions were determined by offset pointing from \ori, NU Ori, or other stars of accurately known position \citep{jw88}, and in a few cases by centering a fainter star. One slit position was chosen to closely match the location of the widely used study of \citet{b91}.The location of the slits settings are shown in Figure \ref{fig:Slits}.  Several systems of designation were employed. The early observations made exclusively for this survey and a comparison of EON radio and optical results \citep{og09} are designated with numbers 1 through 28, while those made specifically for a program of comparison with regions studied with infrared spectroscopy by the Spizer spacecraft (Robert Rubin, Principal Investigator, in preparation) have the names convenient to that program (JW831-M1M2M3, M4, V1, V3, and JW871-SW), and the observations in December, 2009 are designated with upper-case letters.  In many cases the long-slit samples were broken up into smaller samples  designated by position within the slit, as shown in Figure \ref{fig:Slits}. The positions, angular lengths, and related information are given in Appendix A.  When a stellar spectrum was present, it was edited out of the image. On the night of 2009 December 14 four wide-slit 30 second observations of NU~Ori were made.

\subsection{Data Reduction}

The data were processed and calibrated using the common IRAF procedures\footnote{IRAF is distributed by the National Optical Astronomy Observatories, which is operated by the Association of Universities for Research in Astronomy, Inc.\ under cooperative agreement with the National Science foundation.}. As mentioned previously, multiple standard stars were employed to provide a cross-check of the photometric results. 

Because we are interested in some lines that appear both in the nebular and sky spectrum, careful attention was applied to subtraction of the background sky signal. The procedure adopted was to scale a near-time sky spectrum so that its strongest line (the [OI] 5577 \AA\ line so that it matched the signal in that line in the nebular spectrum, then the scaled sky spectrum was subtracted from the nebular spectrum.  This procedure assumes that the sky [OI] 5577 \AA\ line is much stronger than the [OI] 6300 \AA\ + 6363 \AA\ doublet, which one sees is obviously the case from examination of the sky spectra. It also assumes that the nebular [OI] 5577 \AA\ line is much fainter than the [OI] 6300 \AA\ + 6363 \AA\ doublet, which is what is expected from electron temperatures that apply to the Orion Nebula. We are also interested in the [NI] 5197.9  \AA\ + 5200.3 \AA\ doublet, which appears in both spectra, but it is much weaker in the sky spectra than the [OI] 5577 \AA\ line \citep{mf00}. This means that greatest uncertainty in the subtraction procedure is that the relative fluxes in the [OI] and [NI] lines are assumed to remain the same during the sky and nebular spectra. Sodium lamps are used for outdoor lighting in a town near CTIO and the D lines at 5893 \AA\  are variable according to the time of night and evening of the week. This produced some uncertainty in sky subtraction in the shoulder of the HeI 5876 \AA\ line, which should not affect the the accuracy of the photometry of that line by more than a few percent even in the worse case.  A similar condition applies for the weaker city light produced Hg~I 4358 \AA\ line which falls between the strong H$\gamma$ 4340 \AA\ and important temperature diagnostic  [OIII] 4363 \AA\ lines. 

\begin{figure}
\epsscale{1.0}
\plotone{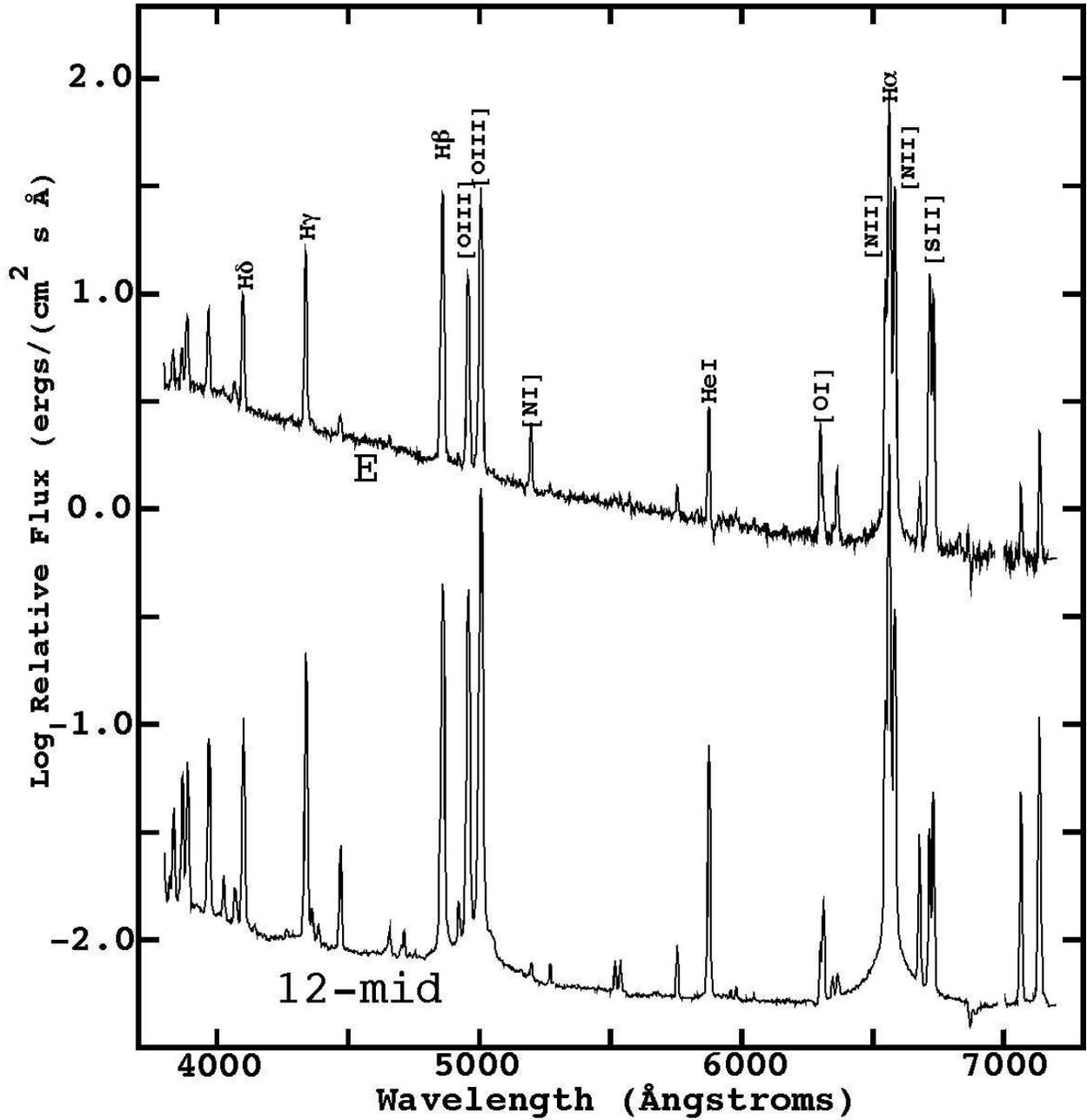}
\caption{The full wavelength range spectrum of the bright 12-mid and faint E samples are shown in relative logarithmic flux plots. A few of the brighter lines are identified. The gap at about 7000 \AA\ is due to a bad column in the CCD detector. The dip long of the 5876 \AA\ HeI line is due to over-subtraction of the NaI lines arising from man-made sky background. These are observed fluxes and have not been reddening corrected.
\label{fig:Spectra}}
\end{figure}

Two examples of calibrated spectra are presented in Figure \ref{fig:Spectra}. These are for the bright 12-mid sample (exposure 120 s) and the intrinsically much fainter E sample (exposure 3000 s). We see from examination of this figure that there were important changes in the spectrum between the Huygens Region, where sample 12-mid is located and the EON, as indicated by sample E. We also see that the longer exposure time for sample E did not produced as high a signal to noise ratio as for sample 12-mid.

Fluxes in the emission-lines were measured using IRAF task splot, which interactively fits a profile to each designated line and produces a total flux in that line.  It was found that Voigt profiles matched the lines best. Splot allows one to deblend multiple lines, which was important in obtaining accurate fluxes of partially resolved lines like the [SII] doublet (6717 \AA\ +6731 \AA) and the [NII] lines (6548 \AA\ + 6583 \AA) that straddle the H$\alpha$ 6563 \AA\ line.  The splot task also provides the relative strength of the underlying continuum in terms of the strength of the targeted emission-line. This is expressed as the equivalent width, i.e. the width of the nebular continuum that will produce the same flux as the emission-line. We recorded this for the \Hbeta\ line and record the strength as the quantity EW (\AA).  Smaller values of EW indicate a continuum relatively stronger to the emission-line. The derived values of the EW for \Hbeta\ are given for each sample in Appendix A. 

Not all emission features were measured. This was because some features were expected to be blends of unresolved lines. The previous high resolution studies of \citet{b00} and \citet{est04} were particularly useful in line identification and selecting which features were most likely to be affected by blends of different ions. In the case of blends arising from the same ion, we often derived individual line fluxes using the deblend feature of splot and then added the components for a total flux.  

\subsection{Correction for Interstellar Extinction}

The amount of extinction across the Huygens Region and the EON is highly variable. The most detailed study is that of \citet{oyz}, who compared high resolution radio continuum images with flux calibrated hydrogen line images to determine point to point values of the extinction at a resolution of a few arcseconds.  \citet{oyz} established that the extinction is greatest to the east of the Trapezium stars, in the so-called Dark Bay region, then rapidly drops to the southwest of the Trapezium stars.  

Appendix A gives the derived values of \chb\ and the extinction corrected surface brightness for each of our samples.  We have adopted the reddening curve recently determined by \citet{bla} from the study of H~I and He~I lines within the Huygens Region (Appendix B) and determined the interstellar extinction from the observed relative flux of the \Halpha\ and \Hbeta\ lines, under the assumption that the intrinsic ratio is 2.89, a value appropriate to the temperature and density conditions for the Orion Nebula \citep{agn3,og09}. For reasons discussed in \S\ 2.3 and \citet{og09}, when the observed flux ratio fell below 2.89 it was assumed that the extinction was zero.  

Extinction corrected relative flux ratios for the spectra are given in Appendix B.  The probable error of the line ratios are determined by the total signal in the sample and the relative strength of the targeted line.  In the innermost samples the uncertainties grow from a few percent to about 20 \%\  at relative line strengths of about 0.1 \%\ of \Hbeta\ and in the fainter outer samples this uncertainty is reached at about 1 \%\ of \Hbeta.

\subsection{Comparison with the earlier study of Baldwin and Collaborators}

The study of \citet{b91} (henceforth B1991) has served as a source of spectrophotometric information used in subsequent investigations. It was also used for calibration of the Hubble Space Telescope WFPC2 and ACS-WFC narrow-band filters \citep{od04,od09}.  B1991used slightly lower wavelength resolution than in the present study, which may affect the deblending of overlapping lines, but the absolute calibration of the isolated \Hbeta\ line should be the same. It is important to compare the results of B1991 and our study.

B1991 used an east-west oriented long slit located west of \ori\ centered on the faint star JW337, which lies about 4\arcsec\ south of \ori. In the original paper it was said that the declination of their slit was the same as \ori, but in private communication Dr. Baldwin communicated the correct pointing. They divided their spectrum into 21 short samples. Since there is some uncertainty of the exact Right Ascension of these samples, the best comparison is to use the larger samples, where the variation of surface brightness is slow. We identified on our  December 2009 spectrum with JW 337 centered, a single sample extending over B1991 samples 10+11+12, where they found an average surface brightness in \Hbeta\ of \Shb=1.73x10$\rm ^{-13}$ \funits\ \arcsecsq. Our derived \Shb\ on that spectrum is 1.90x10$\rm ^{-13}$  \funits\ \arcsecsq, that is, agreement to within 10\%.  A similar comparison was made with the 2008 data and used in the comparison with VLA long wavelength images \citep{og09}. In that study it was assumed that the slit width was 2.6\arcsec, but the comparison with B1991 indicated that the new data needed to be multiplied by 1.21. During the December 2009 observations we discovered that the method of setting the width of the entrance slit used in the November 2008 and January 2009 was incorrect, which would account for the discrepancy if the slit width used earlier had been 1.9\arcsec. Comparison of other regions in the earlier and more recent observing runs led to the same conclusion.  We are satisfied that our absolute calibration using the December 2009 observations is correct and the homogeneity of the instrumental sensitivity functions from three different standard stars argues that the relative flux calibration is also accurate.

In Table \ref{tab:B1991Comparison} we present a comparison of our flux ratios for a few strong lines with the results of B1991. We find excellent agreement for the well isolated lines and the poorest agreement in the group of three lines near \Halpha, which were more poorly resolved in B1991.

\section{Derivation of Electron Temperatures and Densities}

We have determined electron temperatures (\Te) and electron densities (\Ne) using the traditional method of emission-line ratios.  Our wavelength resolution and coverage allowed determination of these conditions in the several ionization layers expected for a blister-type H~II Region \citep{od01}. The [S~II] emission arises from very close to the main-ionization-front, where hydrogen ionization ceases and there we employed the nebular  to auroral reddening corrected line ratios [I(6716 \AA)+I(6731 \AA)]/[I(4069 \AA)+I(4076 \AA)] to determine \Te\ and the nebular doublet ratio I(6716 \AA)/I(6731 \AA) to determine \Ne. The [N~II] emission arises from the adjacent zone with ionized hydrogen but neutral helium and we employed the [I(6583 \AA)+I(6548 \AA)]/I(5755 \AA) ratio to determine \Te. The [Cl~III] emission arises from the outer parts of the ionized hydrogen-neutral helium zone and the inner parts of the next zone where both ions are ionized and we used the doublet ratio I(5518 \AA)/I(5538 \AA) to determine \Ne. [O~III] emission arises from the ionized hydrogen-ionized helium zone and the nebular to auroral ratio [I(5007 \AA)+I(4959)]/I(4363 \AA) was used to determine \Te. We employed the convenient IRAF/STSDAS task ``temden'' \citep{sd94} for our solutions, beginning at the ionization front and working out in an iterative way since there is a slight dependence of the derived \Te\ on the assumed \Ne\ and a slight dependence of the derived \Ne\ on the assumed \Te.  We did not determine [O~III] temperatures when I(4363 \AA) was less than 0.003 or [Cl~III] densities when either line of the doublet was less than 0.004 because of problems in deblending the 4363 \AA\ line from the much stronger H$\gamma$\ 4340 \AA line and discriminating weak [Cl~III] lines against the irregular background continuum. The results of these calculations are presented in Appendix C, Figure \ref{fig:Temps}, and Figure \ref{fig:Densities}.  

We see an indication of a rising \Te\ in the innermost samples for both [N~II] and [S~II].  There are a few anomalously high values of \Te\ from [N~II] and [S~II] for the easternmost samples that fall along the northwest boundary of the nebula in slits 1, 2, 3, 4, 6, and 10. This may be due to selective enhancement of the shorter wavelength lines caused by scattering, as discussed in \S\ 5.1.  The same effect probably accounts for the anomalously high \Te\ values for many of the outer samples.  In the range of distances of from 4\arcmin\ to 8\arcmin, where scattering is not dominant (samples 1-east,16-east,16-mid,16-west,17,18, 20, M4), the average temperatures are 7390$\pm$780 K for ]S~II], 8390$\pm$170 K for [N~II], and 8610$\pm$640 K for [O~III]. This means that there is a small increase in electron temperature as one moves away from the ionization front, the opposite sense of the change expected from radiation hardening (as the optical depth in hydrogen becomes important it selectively removes the lower energy photons, leaving higher energy photons to be absorbed closer to the ionization front, where they will produce more energy input per photoionization).  Robert Rubin determined \Te\ values from the [S~II] lines using up-dated atomic parameters for several representative values of the flux ratios and obtained values about 100 K higher than ``temden'', which indicates that the lower temperatures from [S~II] in this study are probably not due to the method of calculation.  The unexpected difference in the [S~II] \Te\  is probably not due to increased \chb\ since an increase of extinction of A$\rm _{V}$=0.1 would lower the derived electron temperature by only about 200 K.

Other studies of the Huygens Region have sampled different regions, however their results are worth noting. We have adopted the reddening corrected flux ratios from the studies of \citet{deo92}, \citet{b00}, and \citet{est04}, applied ``temden'' and found for \citet{deo92} \Te = 8700, 9700, and 9100 K for [S~II], [N~II], and [O~III] respectively. \citet{b00} fluxes give a wider range of 16500, 10500, and 8400 K while the \citet{est04} flux values give 13800, 10600, and 8400 K. The ``temden'' derived values of \Te\ for the [S~II]  \cite{est04} fluxes are very different  from the value (9050$\pm$800 K) given in the paper and there may be an editing error in their published [S~II] fluxes. None of these single sample studies show the same progression of average temperature that we find in our multiple-samples study.  Our results can be compared with those of two radio studies of the Huygens Region. \citet{wil97} derived \Te = 8300$\pm$200 K from hydrogen 64$\alpha$ to continuum ratios at 42\arcsec\ resolution. More recently \cite{dicker} compared 1.5 GHz archived VLA images with 21.5 GHz archived single-dish Robert C. Byrd Green Bank Telescope data to derive \Te\ from the temperature dependence of the optical depth ratio at this two frequencies, finding \Te =11376$\pm$1050 K for the Huygens Region.  The former radio study involves a more uniform data-set and is not subject to the systematic errors that can creep in when combining data from two very different telescopes and is probably to be preferred.  This means that the \citet{wil97} study, which measures emission from the entire ionized hydrogen zone is in good agreement with the results of this study. A more convolved approach to determining \Te\ is that of \citet{gar08} who compared the line widths of the \Halpha\ and [O~III] 5007 \AA\  lines for the entire Huygens Region. After correction for instrumental and non-thermal broadening components, they derived a thermal broadening corresponding to 9190 K.  In a Space Telescope Imaging Spectrometer study of variations of \Te\ along four long-slit samples in the Huygens Region, \citet{rubin03} found that the [N~II] temperatures were systematically about 1000 K higher than the [O~III] temperatures, in disagreement with our results. That study used multiple grating settings, with the inherent uncertainties in calibration which we tried to circumvent in the present investigation.

We have derived densities from the nebular doublet ratios for [S~II] [I(6716 \AA)/I(6731\AA)] and [Cl~III] [I(5518 \AA)/I(5531 \AA)].  [S~II] was strong in all our spectra and allowed determination of many densities, whereas the higher ionization [Cl~III] emission fades rapidly with distance from \ori\ and many fewer determinations were made.  In this study all of our samples having densities derived from both ions are subject to the scattering discussed in \S\ 5.1. In this case the wavelength dependent scattering means that selectively more of the higher density inner [Cl~III] radiation is scattered, thus giving higher densities than those derived from [S~II].

A \Ne $\sim$ D$\rm ^{-2}$ line is shown in Figure \ref{fig:Densities} for reference and  fits the observed points from about D=2\arcmin\ to 5\arcmin. The observed variation of density and surface brightness potentially allows improving on the 3-D model of \citet{wo95} out to the distance where scattered light begins to be important. At large distances both the observed surface brightness and the derived densities will be larger than the true local values.

In the case of the of M~43 samples electron temperatures could not be derived from [O~III] emission because of the weakness of these lines. The average [S~II] temperature of 7270$\pm$630 K and [N~II] temperature of 7950$\pm$640 K are cooler than for the Huygens region innermost samples, which reflects the cooler color temperature of NU Ori, the exciting star.  Again there is the suggestion that the electron temperature is lower near the local ionization front. The average [S~II] density of 510$\pm$40 \cmq\  of the close-in M~43 samples
is not reached in the Huygens Region until about  3.5\arcmin\ distance from \ori, which probably reflects the lower ionizing flux luminosity of NU Ori. 

 It is perhaps significant that the same unexpected progression of electron temperatures (lower nearer the ionization front) is also found in the well studied and well resolved planetary nebula NGC~6720 \citep{od09b}. This is discussed in \S\ \ref{sect:scattering}.

\section{Position Dependent Diagnostic Line Ratios \label{sect:groups}}

Examination of the dependance of observed properties with increasing distance from the dominant ionizing star (\ori) can be very useful in understanding the 3-D structure of the EON. This is because first-order considerations of photoionization theory can be a guide to interpretation of variations of line ratios and surface brightnesses. Comparison with the expectations of a more elaborate and accurate photoionization model like Cloudy \citep{fer98} would be better, but a simple approach is probably a necessary start.  We have worked with extinction corrected values because these more closely correspond to the nebula itself. However, it must be recalled that extinction corrections derived from the observed F(\Halpha)/F(\Hbeta) ratio will be affected by a varying component of wavelength dependent scattered light and that when this ratio fell below the theoretical value we applied no extinction correction.

We adopted several diagnostic ratios. The first was \Shb. The  second diagnostic ratio was the EW of the continuum with respect to the \Hbeta\ line. Finally, three ratios indicating ionization changes were used, (I([O~III])/I(\Hbeta), I([N~II])/I(\Halpha), and I(([S~II])/I(\Halpha).   The EW is independent of uncertainties in the extinction correction, while the three ionization ratios are insensitive to extinction uncertainties because they reference a hydrogen recombination line at a nearby wavelength. The greatest effect of extinction correction uncertainties will be on \Shb.

The data set of samples were initially divided into subsets called ``groups'' according to their location on an image of the EON. The first natural group were the samples near NU Ori, the dominant star in M~43 and are called ``M43''. The second natural group by location were those samples lying to the northeast of \ori, where the extinction in the foreground veil is the highest, this group is designated as the ``NEArc''. The third natural group (``Outer'') by location were those samples lying along the irregular outer bright Rim of the EON. Like many images of the EON, that used in Figure \ref{fig:Slits} employed broad bandpasses, which don't clearly indicate the distribution of emission-line radiation, yet shows the distribution of scattered stellar continuum radiation quite well. Outer is the largest structure and includes the feature to the southeast of \ori\ called ``North-South Rim''  by \citet{og09}.  The remaining samples fell within the boundary of the EON are were called ``Inner''. We then looked for common values and similar changes with D (the projected distance from \ori) for the non-M~43 samples.  This examination caused some changes in the membership in the groups as ``border'' samples were shifted and led to the breakdown of the Inner group into multiple subgroups.  In some cases individual samples violate one or two of the several criteria for membership within a group and in other cases there is not a clear distinction between the groups, but when there were two or more criteria for membership in a specific group, the sample was placed there. The final membership in groups is also indicated in Figure \ref{fig:Slits}. The process of assigning membership into groups requires integration of multiple parameters, which at some point becomes rather qualitative, much like the identification of constellations in the starry sky, but the surviving groups are probably quite useful in determining the 3-D properties of the EON, as discussed in S\  \ref{sect:3D}.

\section{Discussion}

This large body of new data can be used to determine properties of the Huygens Region, the EON, and the M~43 region.  We first address the 3-D structure of these regions, comment on how this model may be common,  discuss the results we obtain for the regions with known extended X-ray emission, then assess the characteristics of the scattered light and the affect that scattered light has on the interpretation of conditions in the Orion Nebula and similar gaseous nebulae.

\subsection{The 3-D Structure of the EON \label{sect:3D}}

\begin{figure}
\epsscale{1.0}
\plotone{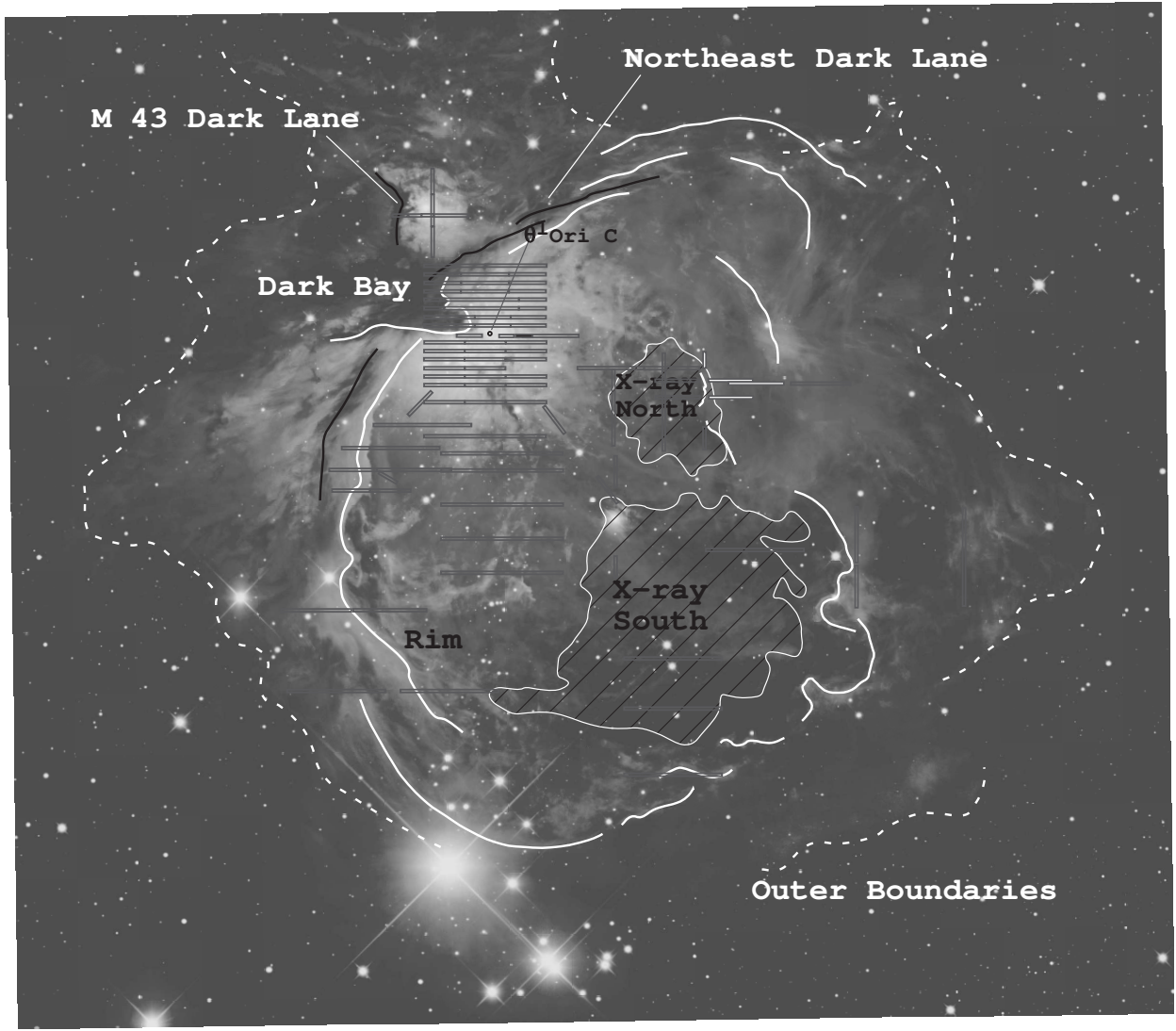}
\caption{The same images as Figure \ref{fig:Slits} except rendered in grayscale. The heavy solid lines trace the boundaries of the EON called the Rim, the heavy dark lines trace high extinction elongated Lanes, dashed lines the outer boundaries of the M~41, M~43, and EON complex, and the light solid lines with internal cross-hatching trace known regions of extended X-ray emission.
\label{fig:sketch}}
\end{figure}

We can use the results from the wide array of samples to determine information about the 3-D structure of the nebula. This is done in a progression of steps, from the most populated groups to those with only a few members. We begin with the Inner group, where scattering of Huygens region light only becomes important in the most distant samples, then consider the M~43 samples, and finally consider the other groups. Figure \ref{fig:sketch} de-emphasizes the positions of the samples and labels the optical features used in the following discussion.

\subsubsection{The Inner Group \label{sect:Inner}}

\begin{figure}
\epsscale{1.0}
\plotone{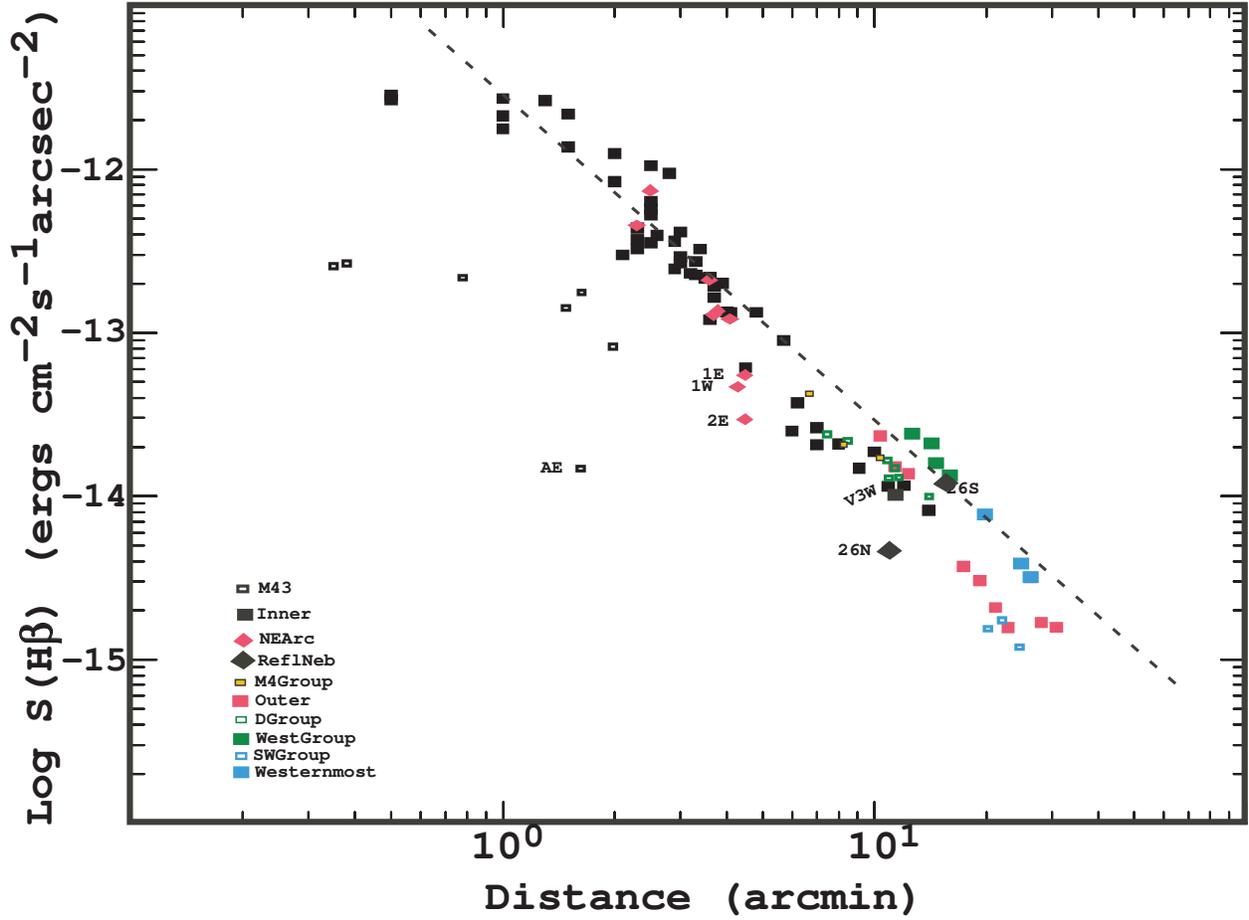}
\caption{The average extinction corrected surface brightness in the \Hbeta\ line for each of our samples are shown as a function of distance from the dominant ionizing star \ori, with the exception of the M43 group samples and they are plotted against the distance from their much cooler exciting star NU Ori.
Logarithmic scales are used for both axes and a \Shb$\sim$D$\rm ^{-2}$ is shown for reference.
We have grouped the samples into subsets shown as the different symbols, these being the same subsets indicated in Figure \ref{fig:Slits}, but the color coding is not always the same.
A few outlying samples are labeled.
\label{fig:DSHb}}
\end{figure}

Examination of Figure \ref{fig:DSHb} shows quantitatively what the eye immediately sees, that the EON progressively grows much fainter further from \ori.
The observed drop is less because the inner samples have had systematically higher corrections for interstellar extinction.  In the smallest values of D there is a flattening which is probably due to the drop with distance being masked by the effects of the sample sizes being comparable to the distance from \ori. Beyond this inner region (essentially the Huygens Region), one sees a steady drop in \Shb.  The overall drop resembles the D$\rm ^{-2}$ relation observed in the region of \ori. In addition to the effects of geometric dilution and a changing tilt of the ionization front there will be removal of ionizing photons by intervening  nebular gas (which would reduce the local \Shb\  \citep{wo95}. Moreover, \citet{og09} argue that scattered light from the Huygens Region begins to be important at about D=5\arcmin. If the albedo of the scattering particles were unity and the phase function isotropic, scattered \Hbeta\ would give a surface brightness relation close to the dashed line in Figure \ref{fig:DSHb}, but of course neither of these conditions are likely to be satisfied.  

The EW is determined by the ratio of incident ionizing LyC photons at the main-ionization-front (which produce the photoionizations that yield \Hbeta) to the scattered stellar continuum. In the Huygens Region \ori\ is the most important source of LyC photons, but as one moves out LyC photons from \oriA\  (spectral type 09.5V \citep{wh77}) may also become important. \oriA\ is difficult to see in Figure \ref{fig:Slits}, but lies south of the boundary between samples 12-east and 12-mid.  We don't know the relative displacement of these two stars along the line-of-sight, but we do know that the giant proplyd 244-440 near \oriA\  indicates by the location of its bright rim that it is dominantly ionized by \oriA\  \citep{jb01b} and that the proplyd 206-447 shows two bright edges \citep{ow94}, indicating ionization by both \oriA\  and \ori.  However, these objects must lie within the open cavity of the nebula and the nebular emission arises mostly from near the main-ionization-front lying further away. The situation with respect to sources of the optical continuum is even more complex. The four brightest Trapezium stars  $\theta \rm ^{1}$Ori A (V=6.72),  $\theta \rm ^{1}$Ori B (V=8.10), \ori\ (V=5.40), and $\theta \rm ^{1}$Ori  D (V=6.70) respectively) dominate the inner Huygens Region, but \oriA\  (V=5.08) and $\theta \rm ^{2}$Ori B (V=6.39) may become important to the southeast.  

\begin{figure}
\epsscale{1.0}
\plotone{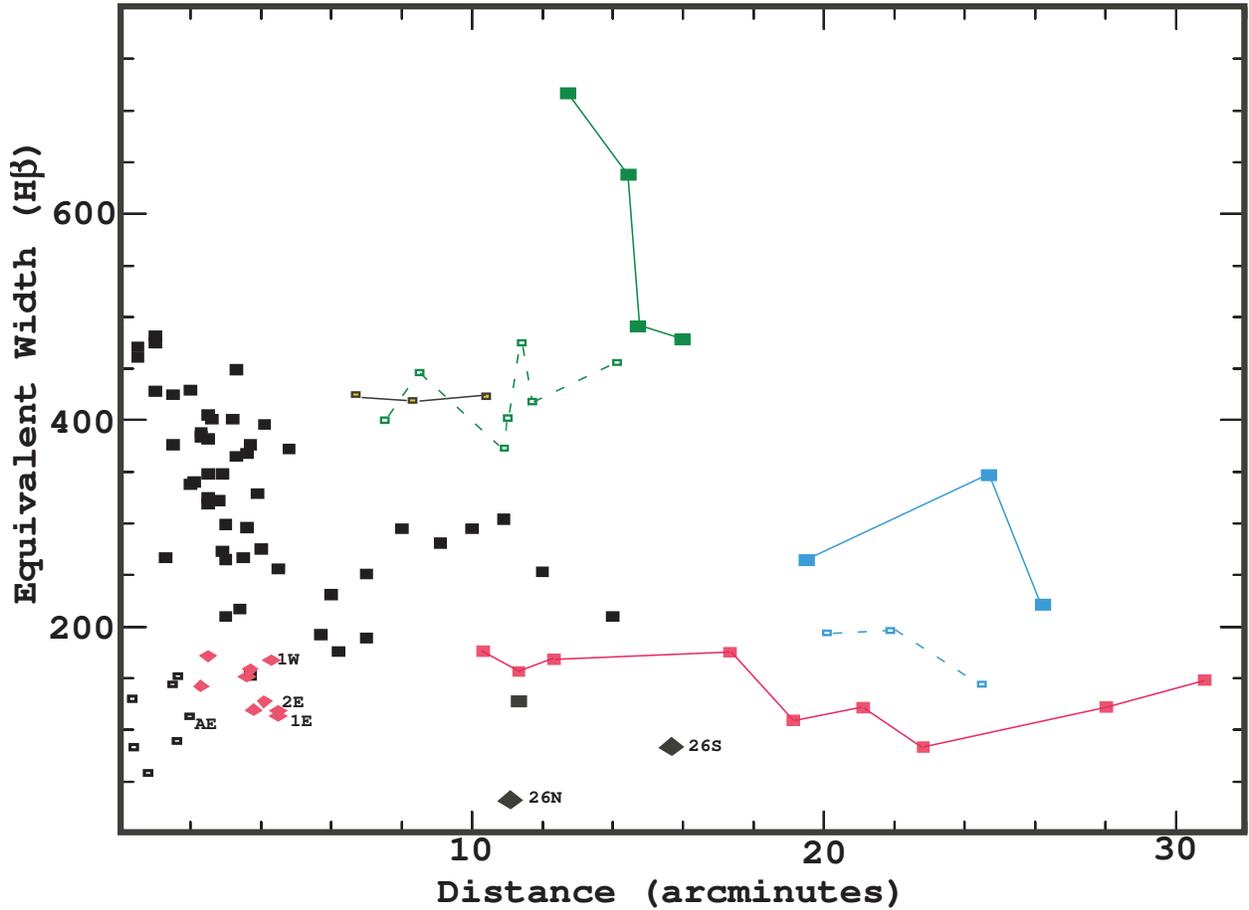}
\caption{The average equivalent widths of the continuum at 4861 \AA\  and the \Hbeta\ emission-line are shown as a function of distance from the dominant ionizing star.
Smaller values of the equivalent width indicate relatively stronger continuums. Since the equivalent width for a purely gaseous nebula at 9000 K is 1800 \AA\ at \Ne =6000 \cmq\ and 1600 \AA\ at \Ne =200 \cmq, most of the continuum arises from scattered starlight.  
The symbols are the same meaning as in Figure \ref{fig:DSHb} and outlying samples have again been labeled. Lines have been drawn between members of some groups for clarity in identifying member samples.
\label{fig:DEW}}
\end{figure}

Under the reasonable assumption that the emission-lines arise from gas ionized by \ori\  on the observer's side of the main-ionization-front and the continuum arises from Trapezium starlight scattered by dust in the dense PDR on the far side of the main-ionization-front, we can predict a general relation between EW and D. The first order expectation is that EW should be constant with D. However, absorption of LyC photons by intervening gas will be greater than the dust extinction of optical photons because of the very high absorption coefficient for hydrogen. As a result, the local ionization front would see relatively fewer ionizing photons that produce the \Hbeta\ emission (as compared with optical photons that produce the scattered continuum) and this leads to the expectation that EW should decrease with distance. This is probably what produces the general drop in the EW values for the Inner group with increasing values of D seen in Figure \ref{fig:DEW}.  

\begin{figure}
\epsscale{1.0}
\plotone{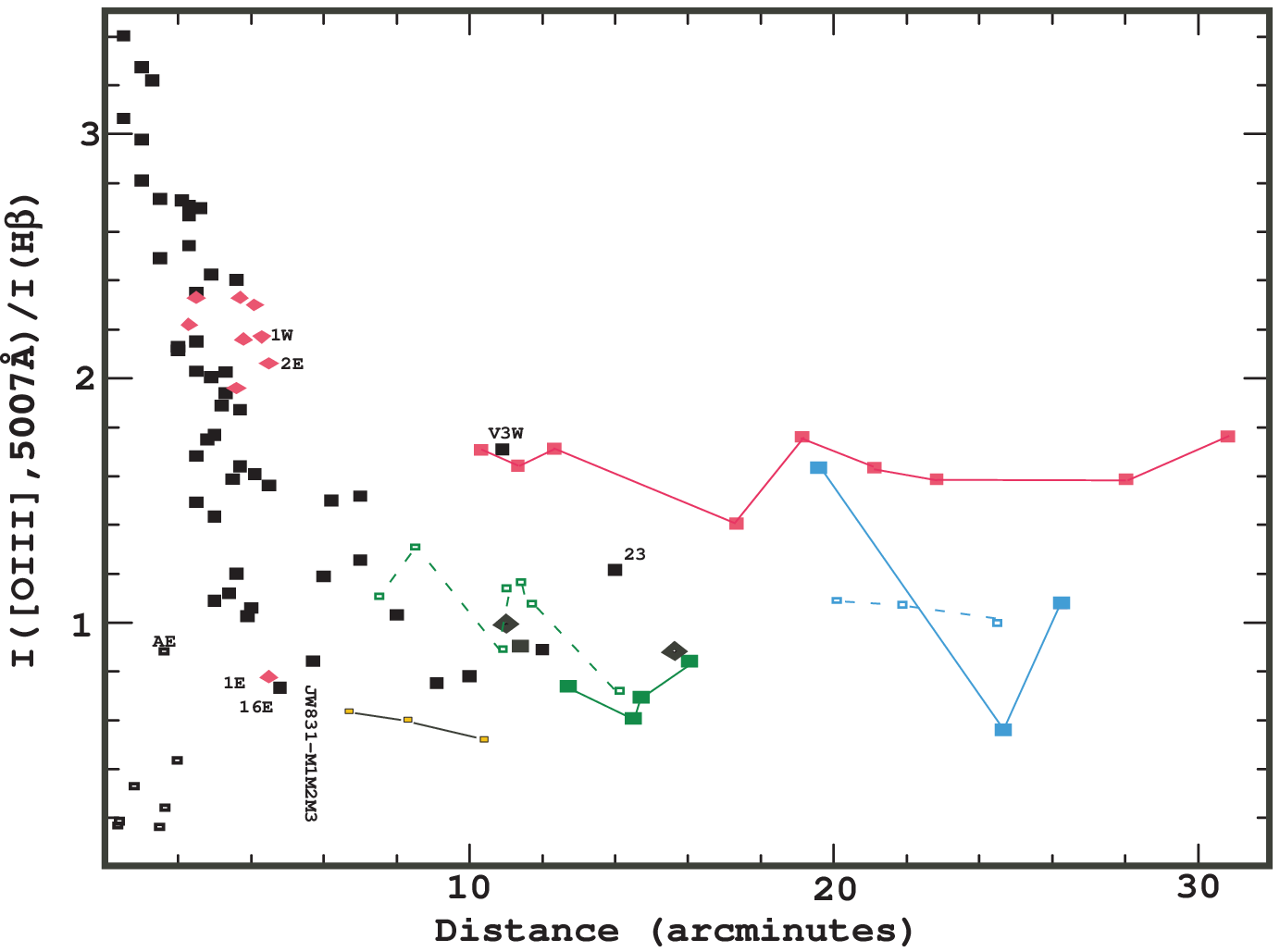}
\caption{The average extinction corrected  I([O~III])/ I(\Hbeta) ratios for each of our samples are shown as a function of distance from the dominant ionizing star.
The symbols are the same meaning as in Figure \ref{fig:DSHb} and a few outlying samples have again been labeled. Lines have been drawn between members of some groups for clarity in identifying member samples.
\label{fig:DOIII}}
\end{figure}

\begin{figure}
\epsscale{1.0}
\plotone{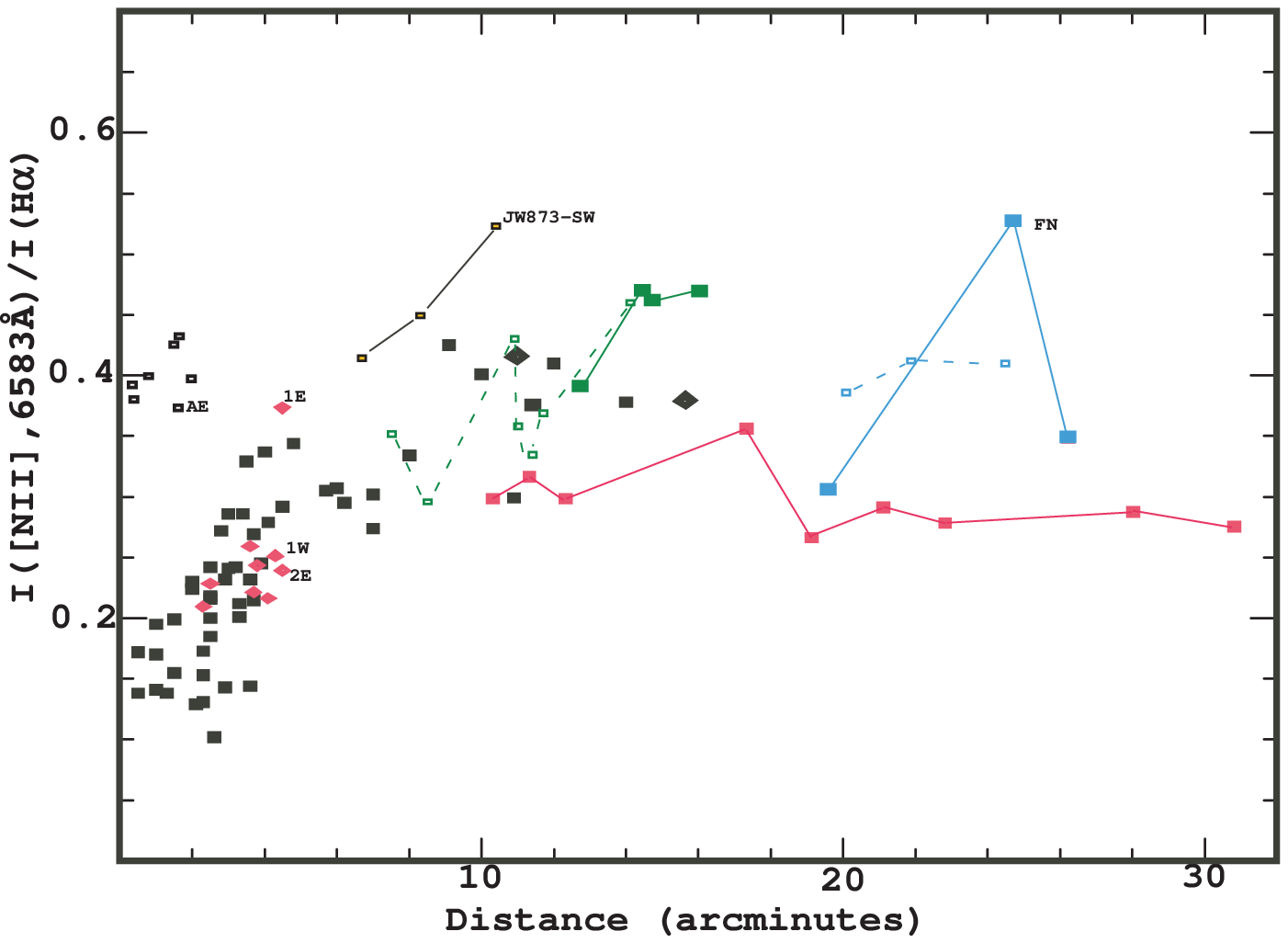}
\caption{The average extinction corrected  I([N~II])/ I(\Halpha) ratios for each of our samples are shown as a function of distance from the dominant ionizing star.
The symbols are the same meaning as in Figure \ref{fig:DSHb} and a few outlying samples have again been labeled. Lines have been drawn between members of some groups for clarity in identifying member samples.
\label{fig:DNII}}
\end{figure}

\begin{figure}
\epsscale{1.0}
\plotone{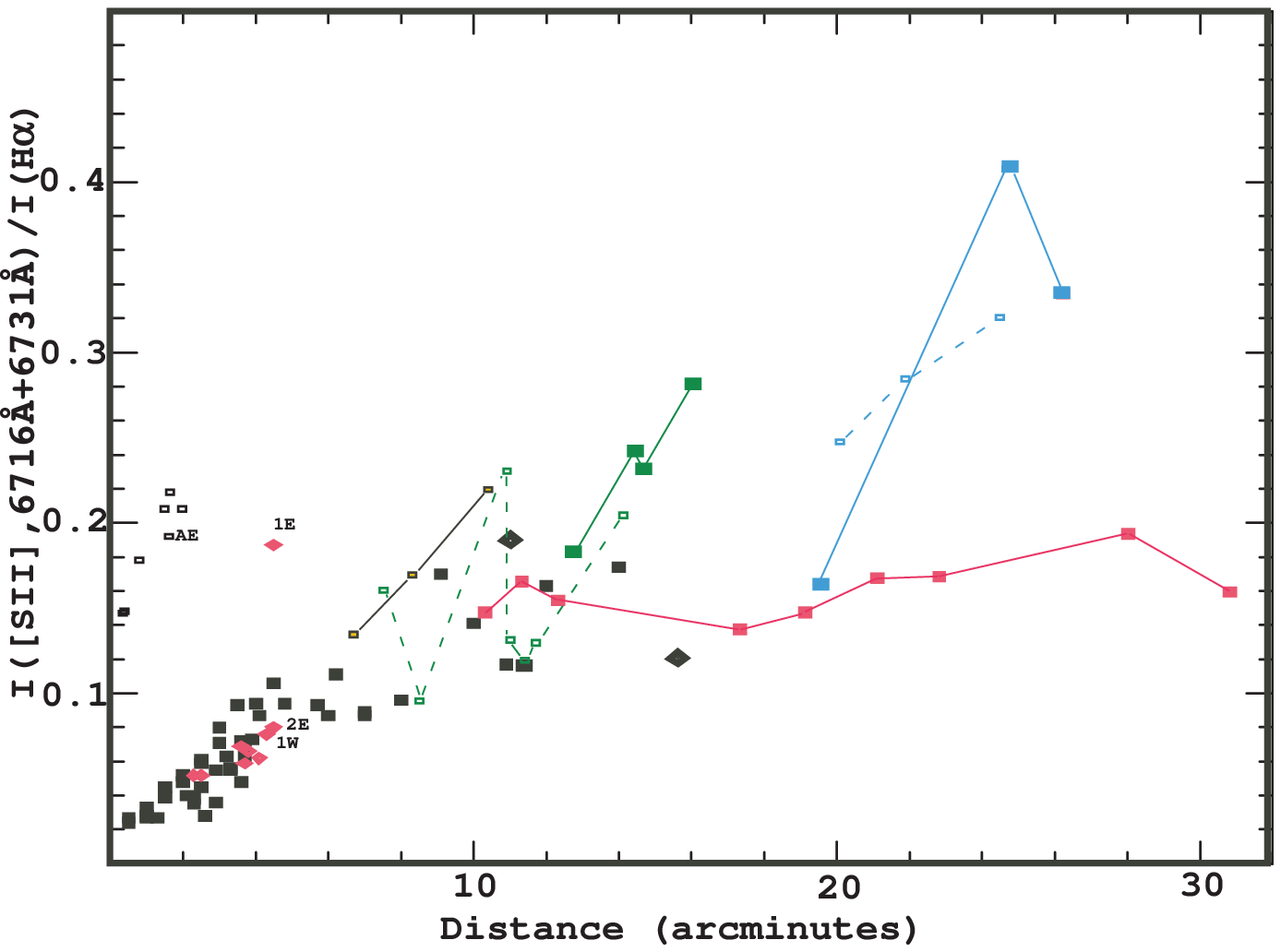}
\caption{The average extinction corrected  I([S~II])/ I(\Halpha) ratios for each of our samples are shown as a function of distance from the dominant ionizing star.
The symbols are the same meaning as in Figure \ref{fig:DSHb} and a few outlying samples have  been labeled. Lines have been drawn between members of some groups for clarity in identifying member samples.
\label{fig:DSII}}
\end{figure}

When assessing the general trends for the ionic ratios we have to recall the general properties of the emission as described in paragraph 3 of \S\ 1.
[O~III] arises from the region closest to the ionizing star, where both hydrogen and helium are ionized, [N~II] arises from the region where hydrogen is ionized but helium is neutral, and [S~II] arises from very near the main-ionization-front where there is a delicate balance between hydrogen ionizing photons (which can remove through photoionization the S$^{+}$ that produces [S~II] emission) and the electrons arising from ionization of hydrogen that cause the collisional excitation of [S~II]. The Balmer lines arise from throughout the ionized hydrogen zone.  In Figure \ref{fig:DOIII} we see that I([O~III])/I(\Hbeta) drops with increasing distance for the Inner samples,  which indicates that photoionization of helium is more complete in the inner regions while there are fewer ionizing photons further out and a correspondingly smaller [O~III] emitting zone.  The increase in I([N~II])/I(\Halpha)  (Figure \ref{fig:DNII}) and I([S~II])/I([\Halpha)  (Figure \ref{fig:DSII}) with increasing distance indicates that the [N~II] emitting zone is becoming a bigger fraction of the total ionized hydrogen zone and that the transition zone between neutral and ionized hydrogen (where [S~II] is produced) is also becoming relatively more important.

The innermost of the Inner group  cover the region modeled in detail by \citet{wo95}. Although they did not consider ionization changes expected for their irregular concave model, it appears to be fully compatible with the Huygens Region samples of the Inner group.

\subsubsection{The SW Group \label{sect:SWgroup}}

The positions of these three samples  are clearly within the Rim and were selected in order to determine if the darkest regions of the EON that were inside the Rim feature were different from the other  EON samples. In terms of \Shb\ they fall on an extension of the pattern established with the Inner Group, that is, much lower than a D$\rm ^{-2}$ projection from the Huygens Region and slightly fainter than the Outer Group at similar large distances.  The [O~III]/\Hbeta\ ratio is slightly higher than the lowest values of the Inner Group. The [N~II]/\Halpha\ ratio is lower than a projection of the pattern for the Inner Group. These two line ratios argue that the SW Group is similar to the outer Inner Group except that it has a greater component of Huygens Region scattered light.  The [S~II]/\Halpha\ ratio is only slightly lower than a projection of the Inner Group pattern, which would indicate that the intrinsic [S~II]/\Halpha\ ratio was very strong prior to dilution by Huygens Region scattered light.  EW values are larger than the Outer Group samples of similar distance and slightly smaller than the outer Inner Group.  

The best 3-D interpretation of the SW Group is that this region is very similar to the outer Inner Group and simply carries the patterns established there to greater distances from \ori. The fact that the EW values are intermediate between the outer Inner Group and the Outer samples indicates that the ionization front here is curved towards the observer, but does not lie as far in the foreground as the Rim sampled by the Outer Group.

\subsubsection{The D Group\label{Dgroup}}
The D Group samples are the second set that lie inside the Rim. This region was well sampled because of wanting to supply observation useful for studying the well-defined, west facing shock  that is drawn in Figure \ref{fig:sketch}. This shock was the subject of an early study \citep{rjd} and is the subject of a more comprehensive investigation by W. H. Henney and the observational data on it from this program  will be discussed in his later paper. It is well illustrated in \citet{wjh}. 

The \Shb\  of this group is similar to Inner Group samples at the same distance, averaging only slightly higher. The diagnostic ratios [O~III]/\Hbeta, [N~II]/\Halpha, and [S~II]/\Halpha\ are all essentially identical to the Inner Group at the same distance, where Huygens Region scattered light has begun to be quite important. The D Group does differ significantly in their EW values, being much higher than the Inner Group samples at this distance. Again, this is probably due to less LyC attenuation in this direction. The most straightforward explanation of that is that the material in the D Group samples lie in lower density material displaced towards the observer with respect to the main-ionization-front. In this explanation the even greater EW values of the West Group (\S\ 5.1.8) would be due to the latter's being even closer to the observer.

\subsubsection{The M4 Group}

The M4 Group roughly isolates the samples on a broad, bright, north-south feature lying parallel to the east boundary of the Rim. Probably the best published image of it is in \citet{wjh}.  It is bounded on both sides by regions of lower surface brightness than the general area, which immediately argues that it is a separate feature.  The three samples are similar in \Shb\ to the Inner group at this distance and all of the samples have little extinction.  
However, the M4 group differs in the ionization ratios, with [O~III]/\Hbeta, [N~II]/\Halpha, and [S~II]/\Halpha\ lying on a projection of the pattern of change towards lower ionization established in the inner samples of the Inner group (before Huygens Region scattered light has begun to be important). In addition, the EW values average 423$\pm$6 \AA, which is similar to the samples closest to \ori.  The Inner group samples at these distances have artificially enhanced \Shb\ through scattering of Huygens Region  \Hbeta\ photons, so that the M4 group members are intrinsically much brighter than the nearby Inner group samples. All of these factors argue that the feature is distinctive and the ionizing LyC radiation has been little attenuated before reaching it. The simplest explanation is that it is a feature further from the main-ionization-front and its LyC radiation has passed only through the low density gas found at greater distances from that front. The group does lie in the vicinity of the very large HH~400 outflow \citep{jb01a}, but the similarity  to the main-ionization-front in ionization argues that little of the radiation comes from that shock.

\subsubsection{The ReflNeb Group}

We used one slit that passed through the bright star Parenago 1605 (P1605, V372 Ori) and extracted two samples (26-north and 26-south) that were well removed from contributions from the star. P1605 is an HAEBE variable and point X-ray source of spectral type B9.5. The characteristics of the star are summarized in \citet{ham} and the heliocentric radial velocity of + 49 \kms\ was derived by \citet{hmj}. Johnson's velocity is about 23 \kms\ more positive than the group velocity of other member stars and he states that the velocity is driven by the limited sampling of a binary star system.  The best published image \citep{wjh} shows that the star is surrounded by a reflection nebula, clearly indicating its association with nebular material.  However, the samples taken were outside of the apparent reflection nebula and the effects of instrumental scattering of the starlight. The star is too cool to contribute to ionization of nearby gas. The samples are very similar to the outer Inner Group and D Group at similar distances, but 26-north is much lower in \Shb\  than 26-south and members of the other two groups.  Where the ReflNeb Group differs most from the other groups is in the extraordinarily low values of EW, being 30 \AA\  for 26-north and 80 \AA\  for 26-S. The reason for the low EW values is undoubtedly due to scattering of stellar continuum from P1605 and these two samples become the best representation of the scattering of continuum from a star not associated with the Trapezium. 

\subsubsection{The Samples Associated with M~43}

The M43 group is quite different from the Inner group discussed above, which reflects the fact that NU Ori is much cooler than \ori. The extinction corrected \Shb\  is about a factor of 15 lower than similar distance samples in the Inner group.  This is just about the same as the ratio of LyC luminosities of \ori\ and NU Ori  given in \citet{agn3}. The agreement indicates that the displacement of the ionizing star and the ionization front it causes on the Orion Molecular Cloud is about the same as for the Inner group. The EW values are very small, indicating that the continuum is very strong relative to the \Hbeta\ emission-line, reflecting its similarity to the outer parts of the Inner group, but having even lower intrinsic ionization. The extraordinarily low I([O~III])/I(\Hbeta) ratio  but high I([N~II])/I(\Halpha) and I([S~II])/I(\Halpha) ratios are the results of the much cooler temperature of NU Ori. In fact, at B 0.5 V and 32000 K, the star is right on the usual limit between classification as a reflection nebula or a Diffuse Galactic Nebula. Our results are in agreement with  \citet{peicos69} whose spectroscopic study of a single long-slit sample in M~43 did not go faint enough to detect the He~I 5876 \AA\ line. They used their spectrum to determine that M~43 was not illuminated by ionizing radiation from \ori\ and does not include substantial local scattering of Huygens Region emission-lines.

Only the sample  A-east (AE in the figures)  is not like the others in some of the diagnostics. This sample lies on a north-south dark lane east of NU Ori (Figure \ref{fig:sketch}, where it is labeled M~43 Dark Lane). This dark lane appears to be a foreground feature.  However at  \chb=0 .68,  sample A-east is less reddened than many of the other M43 group samples.  Its \Shb\  is anomalously low for its distance from NU Ori and its [O~III]/\Hbeta\ ratio anomalously  high.  The most straightforward interpretation of these anomalies is that the dark lane is so very optically thick that much of the radiation arises not from the blister portion of M~43 but from ionized material on the observer's side of the dark lane. This would cause the extinction to be underestimated and the extinction-corrected \Shb\  to be too small; however, the large [O~III]/\Hbeta\ ratio suggests a role for photoionization by \ori\ rather than simply an underestimate of the extinction. The distance of sample A-east is 8.7\arcmin\ from \ori. The \Shb\ and [O~III]/\Hbeta\ are about that seen in Inner region samples at this distance.The EW  and [S~II]/\Halpha\ values match a projection of the Inner samples out to 8.7\arcmin\ and [N~II]/\Halpha\ is about the same as Inner samples at this distance from \ori. All of these characteristics indicate that sample A-east is dominated by \ori, rather than NU Ori.  This would indicate that the north-south dark lane does not lie very much closer to the observer than \ori, in contrast with the Dark Bay feature which is almost certainly part of the foreground Veil at about one to three parsecs in the foreground \citep{abel,od09a}.

\subsubsection{The Outer Group}

The Outer group covers the widest area of the EON and is composed of samples along and outside of the irregular bright Rim that defines the EON.
In the \Shb\ figure we see that the samples lie along a projection of the decrease of surface brightness with increasing distance defined by the outer samples of the Inner group. However, we saw in \S\ \ref{sect:Inner} that those outer samples were strongly affected by scattered Huygens Region radiation, which means that the intrinsic surface brightness of those outer members of the Inner group is lower than presented. We show below that this is also true of our Outer group samples.

In the [O~III]/\Hbeta\ figure we see that this ratio is almost constant at a value characteristic of the Inner group at about 3.5\arcmin\ distance and in contrast with the rapid drop with distance in the Inner group. Both the [N~II]/\Halpha\ and [S~II]/\Halpha\ ratios show a similar behavior, in that they are nearly constant at values similar to samples much closer to \ori. These three trends almost certainly indicate that the Outer group radiation is primarily light scattered by a neutral cloud of dusty material and originates little radiation of its own.  The jump in the  [O~III]/\Hbeta\ ratio indicates an abrupt difference in conditions from the outer samples of the Inner Group and that the fraction of scattered Huygens Region light has become dominant. This means that the intrinsic emission in all lines, including \Hbeta, is low and that the true surface brightness is much lower than it appears to be. The eastern parts of the Rim are sharply bounded, so that we must be looking at a low-level ionization front almost edge-on, which would raise the observed surface brightness. The fact that the observed but scattering-corrected surface brightness is low indicates that the Rim regions sampled lie at larger radial distances from from the \ori\ than indicated by their separation in the plane of the sky;  that is, they lie in the foreground.  The Veil must lie further towards the observer since the Dark Bay feature appears to obscure the Rim as it passes to the northeast of \ori.  

The one diagnostic parameter that is markedly different from the inner Huygens Region is the EW, which is nearly constant at a value lower than encountered in the Inner group (with only one exception).  This is consistent with the interpretation of the Rim as being primarily scattered light from the Huygens Region with the addition of continuum from the luminous stars concentrated in the Huygens Region. These stars, including the $\rm \theta ^{2}$ group, must lie in the foreground of the main-ionization-front and are closer to the Rim material, which would account for the stronger continuum in the Outer Group.  

\subsubsection{The Westernmost  and West Groups\label{sect:wwgroups}}

The western side of the EON is much less well-defined than the east and there are a series of features in various emission-line and infrared continuum images that resemble the the east portions of the Rim, but lie at several distances from the center of the EON.  We grouped samples in these remaining areas that by position fall outside of the EON proper, but do not share the properties of the Outer Group. 

Both the Westernmost Group and West Group samples  lie on a D$\rm ^{-2}$ projection from the Huygens Region of the \Hbeta\ surface brightness and are about a factor of two brighter than the Inner  and Outer samples at the same distance. In this sense they look like the same natural group. However, in  some of their other diagnostic parameters they appear to represent different conditions. They have similar [O~III]/\Hbeta\ values (with a wider spread in the Westernmost Group) and these are about like the outer Inner Group samples and much lower than the Outer Group samples. In both groups their [S~II]/\Halpha\ ratios  lie on a projection of the pattern of the Inner sample and are higher than the similar distance Outer samples. The [N~II]/\Halpha\ ratios behave in the same fashion with the exception that the Westernmost Group samples lie a bit lower than the trend established by the Inner Group samples, but still lie higher than the Outer Group samples. 

These diagnostic line ratios indicate that there is less influence of scattered Huygens Region radiation. This information indicates that the regions are ionization fronts and the higher \Shb\  is probably due to those ionization fronts being illuminated by LyC radiation little attenuated by gas in the Huygens Region. Where these two groups differ is that the EW is very high in the West Group (even larger than the central parts of the Huygens Region), while the EW for the Westernmost Group is comparable to parts of the Inner Group just before where scattered Huygens Region radiation begins to be important. However, the EW for the Westernmost Group is still more than twice the value of the Outer Group of similar distance. The difference in the EW of the two groups is probably due to additional absorption of LyC radiation at the greater distances to the Westernmost Group. Their PDR must not be significantly closer to the the stars that contribute the strong scattered continuum in the Outer Group.

\subsubsection{The NEArc Samples}

The NEArc group are those samples taken in the region in and near the Dark Bay feature. In the neutral hydrogen study of \citet{vdw89} this is the region of highest column density of H~I.  The majority of the samples lie among the Inner samples in the D-S(\Hbeta) plot but 1-east, 1-west, and 2-east bunch lower. Except for sample 1-east , the NEArc samples indicate higher ionization than Inner samples at the same distance from \ori.  In the D-EW plot all the NEArc group  cluster near values of EW  one-half that of Inner samples at the same distance.  These facts argue that the NEArc samples are both local emission and scattered light from the Huygens Region; but they are also illuminated by NU Ori, which would increase the scattered light continuum strength without increasing the ionization.  

Sample 1-east is different from the other NEArc samples in that it is of lower ionization than the others. It is also the closest sample to NU Ori and it is likely that this region is sufficiently blocked from radiation from \ori\   and the Huygens Region that it is equally affected by \ori\ and NU Ori.

\subsubsection{Relation of Optical Results to Observations of Extended X-ray Regions \label{sect:xray}}

In an important recent paper \citet{gud} report the discovery of million-degree gas in two extended X-ray emitting regions within the EON. The outlines of these are shown in Figure \ref{fig:sketch}.  By plan, several of our slit samples fall within these X-ray regions with the X-ray North cloud including samples 27-west, C, D, D-Northeast,  D-Southeast, and D-south and the X-ray South cloud including samples 26-south, E, H, and I. We have tried to determine if there are unusual characteristics of the optical radiation from these X-ray clouds. 

Figure \ref{fig:Temps} shows the electron temperatures for these clouds in bold line symbols (the X-ray North samples are the grouping with smaller D values). We can see no detectable different from other samples at similar distances. The similarity is not surprising since most of the radiation at those distances is scattered Huygens Region radiation, rather than being locally emitted. The electron temperatures of material influenced by the X-ray clouds would have to be very different in order to show up in this diagram. In the same fashion, the diagnostic ratios are most simply explained as emission from a low electron temperature gas, as addressed in \S\ \ref{sect:wwgroups}. 

\subsubsection{A Summary of the 3-D Properties of the Different Groups\label{sect:summ3D}}

Understanding the 3-D structure of the Orion Nebula has been the target of multiple investigations for more than half a century.  The assumption of spherical symmetry was the start of many studies but this broke down with the recognition that the increasing blue-shift of emission-lines coming from progressively higher states of ionization indicates that most of the Huygens Region radiation is coming from an ionized blister, beyond the dominant ionizing star (\ori), and on the side that faces the observer of a giant molecular cloud \citep{kaler67,z73,bal74}.  A more quantitative approach, using the method suggested by Ferland \citep{b91},  was adopted by \citet{wo95} to produce an irregular concave model.  As noted in \S\ \ref{sect:intro}, this surface has a nearly edge-on escarpment that passes from northeast to southwest, just west of \oriA\  and leads to the feature called the Bright Bar.  In the foreground is the Veil of essentially neutral material \citep{vdw89} that has the greatest column density in the direction of the Dark Bay.  The distance of the Veil has been variously estimated as being 1 to 3 parsecs in the foreground \citep{abel,od09a}. We now know \citep{od09a} that the bright rise to the southwest of the Trapezium is actually a dense cloud lying in front of the main-ionization-front and is \citep{od09a} the source (called Orion-S) of recent star formation and continued bipolar outflows. Most recently \citet{od09a} have shown that the regions close to \ori\ are affected by the high velocity stellar wind from that star, but that the surrounding shock is open to the southwest. The immediate questions are "What is the structure of the EON and how does this relate to the Huygens Region structure?"
 
The EON is probably a very low density region with little material on the observer's side in the foreground. A significant fraction of this escaping low density gas is heated sufficiently to create extended X-ray  emitting clouds \citep{gud}. The far side of it is an ionized layer and beyond that a PDR of sufficient optical depth to scatter Huygens Region radiation (Inner group, ReflNeb group, D group, SWgroup).  To directly see the Huygens Region, the ionized side of the EON must be concave.  
Within the boundary of the EON there are other features that are only loosely related to the main-ionization-front, for example the M4 group.
The concavity apparently increases with distance and eventually causes the apparent boundary that is called here the Rim (Outer group, West group, Westernmost group) .  The presence of many well defined structures along the rim argue that the EON has begun to curve back towards the center.  

The most difficult things to model in 3-D are the relative locations of the Dark Bay, the Northeast Dark Lane, the M~43 Dark Lane, and M~43 itself with respect to \ori.  We know that M~43 is not illuminated by \ori\ but that the very optically thick Northeast Dark Lane is affected by both \ori\ and NU Ori.  We also know that the M~43 Dark Lane sample is not illuminated by the nearby NU Ori, rather, it is illuminated by \ori.
 
The Rim is much less sharply delineated from position angles 225\arcdeg -315\arcdeg. In this quadrant there are apparent multiple components to the Rim and they are more irregular, some even approaching the shapes of the elephant trunks seen in many other Diffuse Galactic Nebulae.  The fine-scale structure there may be related to the fact that the high velocity wind from \ori\ directly flows into that direction. An additional affect may the LyC shadow-zone cast by the Orion-S imbedded cloud, which will block radiation over a position angle range of about 220\arcdeg\ to 250\arcdeg\ and there are numerous structures with the Rim in those directions.  To the northwest there are features that look like large shocks and these are in approximate alignment with a major outflow from Orion-S \citep{wjh}, suggesting that these structures are shaped by that collimated outflow.  
 
\begin{figure}
\epsscale{1.0}
\plotone{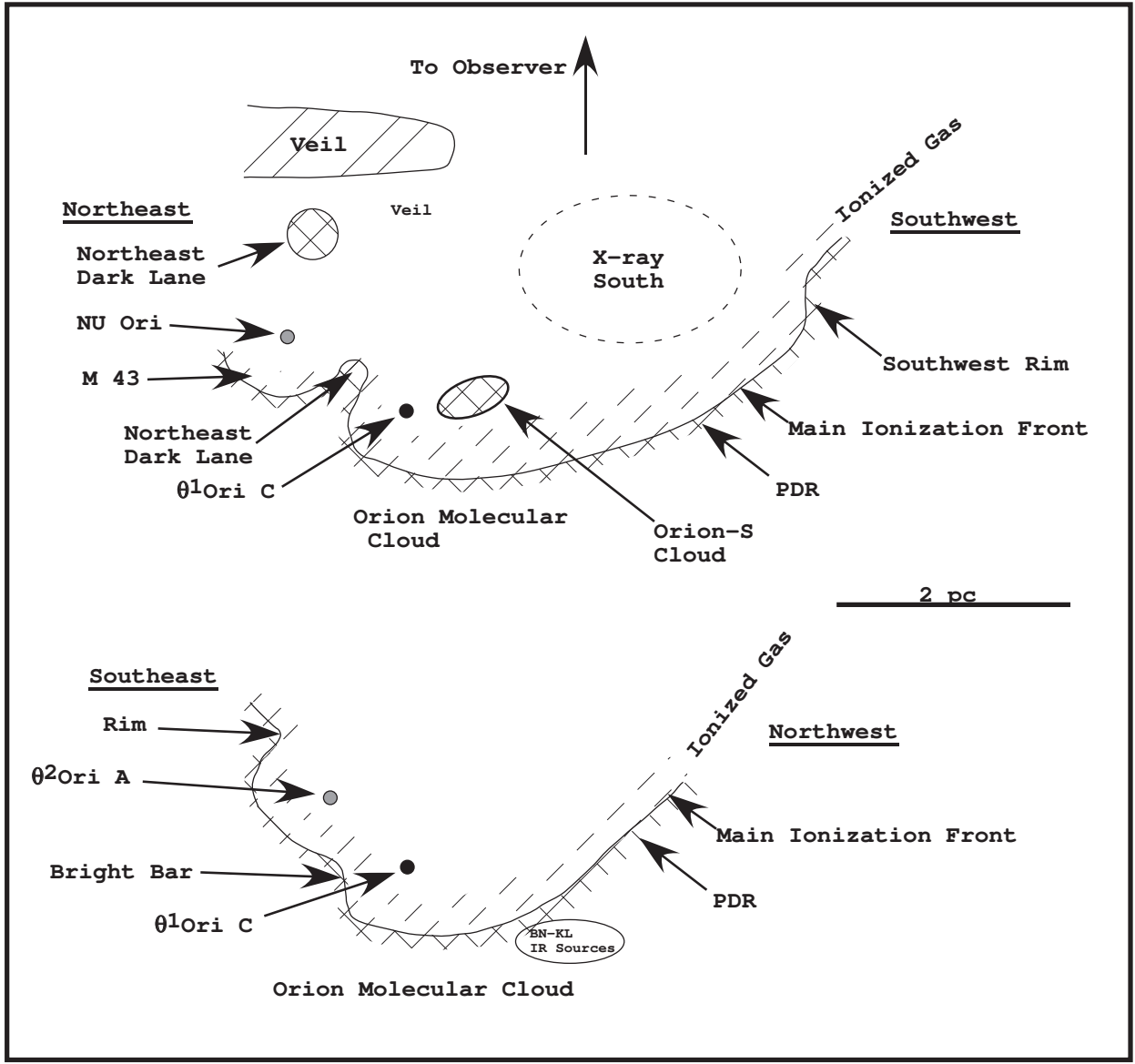}
\caption{Two cross-section cuts perpendicular to the plane of the sky and through the M~42, EON, and M~43 region are shown. The upper panel is from northeast to southwest and passes through both NU Ori and \ori\ while the lower panel is from southeast to northwest and passes through \oriA\ and \ori. The features and details are discussed in \S\ \ref{sect:summ3D}.
\label{fig:cartoon}}
\end{figure}

 We present in Figure \ref{fig:cartoon} two cross-section cuts perpendicular to the plane of the sky and passing through nearly orthogonal angles on the sky. This figure depicts the major features of our suggested structure of the M~42, EON, and M~43 region. It should be noted \citep{od01} that the ionized blister that faces the ionizing star is actually of greatest density near the main-ionization-front and decreases in density with increasing distance from that layer. Likewise, the peak emissivity of the blister decreases with increasing distance from the ionizing star. The northeast-southwest plane was broadened slightly to include the Northeast Dark Lane and Orion-S \citep{od09a} features, which do not fall exactly on the plane including \ori\ and NU Ori. 
 
In a Fabry-Perot study,  \citet{deh73} noted a velocity splitting of the [N~II] 6583 \AA\ line in three regions of the inner EON. Where their more limited coverage overlapped with DeHarveng's  field \citet{gar07} also found line splitting in [S~II] but not [O~I] or [S~III]  and called it the Diffuse Blue Layer. They derived a density of \Ne = 400 \cmq. DeHarveng's region A to the southeast of \ori\ lies just inside the southeast Rim. It is likely that the two velocity components arise from either side of the Rim as it curves transitioning from concave, to convex, and back to concave as shown in the lower panel of Figure \ref{fig:cartoon}. Her region B lies just inside of of the northern portion of the Rim and inside of the Northeast Dark Lane and could arise from emission from both sides of either features. Her region C was not included in our slit samples but does extend to the northwest where the EON has several large arcs.
 
 \subsection{Are Other H~II Regions Similar to the EON?}

In the preceding section we have established a detailed structure for the M~42, EON, and M~43 regions. This was possible because of the dominance of the LyC luminosity by one star (\ori) and the many results obtained from a rich variety of methods of observations (X-rays through radio waves).  It can be argued that building such a detailed structure goes beyond the justification that ``It can be done.'' The factors that determine the structure of the Orion Nebula are also at play in other  Diffuse Galactic Nebulae (H~II Regions) and many would probably resemble the Orion Nebula if examined in similar detail. 

The immediate temptation in interpreting  H~II Regions is to assume that circular symmetry on the plane of the sky argues for spherical symmetry in 3-D. This bias is goes back to the paradigm-establishing paper of \citet{strom}, who first applied photoionization theory  to a star in a constant density cloud of neutral hydrogen. However, we know today \citep{od99} that objects of the same first-order appearance can also be produced by a blister nebula and it can be argued that these are more common. When an H~II region is formed near the edge of a GMC,  the ionized gas is likely to break out of the overlying material (the so-called Champagne Phase \citep{tentag}) and photoionized gas rapidly escapes. This quickly produces a blister nebula that can have a long life. Blisters that are not facing the observer will not be seen visually, so that selectively we will see the blister nebulae nearly face-on. The blister will probably be irregular in form as it will erode more slowly where the underlying density of the GMC is higher \citep{hen05}.

Of course the situation becomes more complex when multiple stars are formed close together and when there are multiple epochs of massive star formation.  However, the basic paradigm and observational restrictions will apply, which means that the Orion Nebula becomes a good exemplar of optically bright H~II Regions. 
This also means that it is a test-bed for evaluating our knowledge of the physics that occurs in H~II Regions \citep{od01} and for determining how accurately we can calculate basic parameters like electron temperatures, electron densities, and abundances.

\subsection{Scattering of Emission-Line Radiation by the PDR\label{sect:scattering}}

There are several lines of evidence that emission-line radiation is scattered in the PDR behind the main-ionization-front . This effect appears to be wavelength dependent and has important repercussions in the interpretation of emission-lines from the Orion Nebula and probably also other  gaseous nebulae.

\subsubsection{Comparison of Radio and Optical Surface Brightness}

The present study extends the discussion of the possible role of Huygens Region light scattered by material in the EON that was initiated in \citet{og09}. In that earlier study a spectroscopic slice of the EON about 8\arcmin\ south of  \ori\ was  compared with a 327.5 MHz low resolution (78.6\arcsec x 65.0\arcsec) image made with the Very Large Array radio telescope. It was concluded that the optically thin radio continuum became systematically fainter with distance from \ori\ than expected from the \Shb. This was interpreted as some of the EON \Hbeta\ optical light having arisen from the Huygens Region and was scattered locally by dust in the high density PDR that lies immediately beyond the ionized layers of the EON.  

\begin{figure}
\epsscale{1.0} 
\plotone{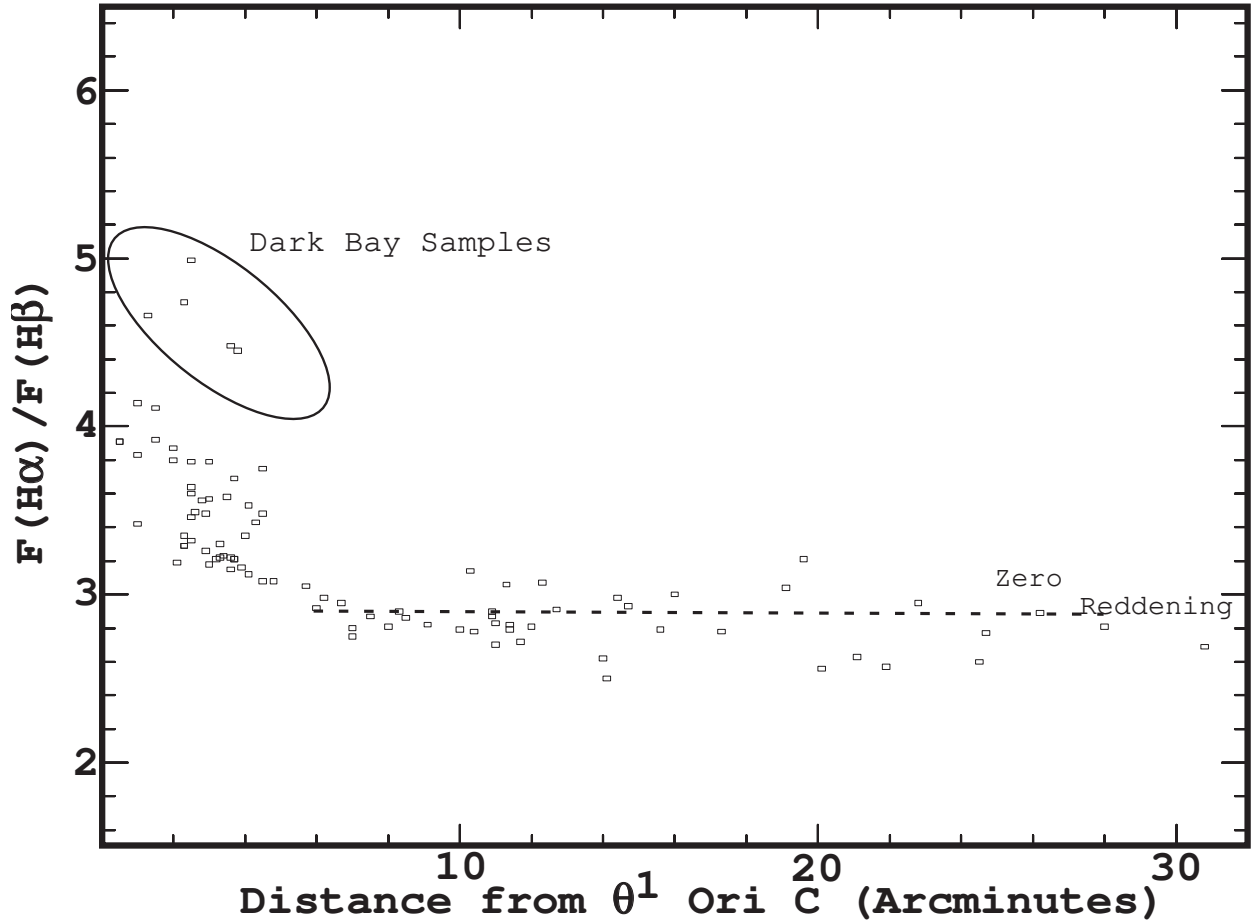}
\caption{This depiction of the observed F(\Halpha)/F(\Hbeta) shows the systematic decrease with increasing distance from \ori. The ellipse shows the five samples (5-East, 6-East, 7-East, 8-East, 9-East) that fall in the Dark Bay feature to the east of the Trapezium. The theoretically expected zero reddening value of 2.89 is also shown.
\label{fig:HaHb}}
\end{figure}

\citet{og09} gave supporting evidence for this interpretation by showing that their new spectroscopic determination of the ratio of fluxes of the first two members of the Balmer series F(\Halpha)/F(\Hbeta)  dropped with increasing distance from \ori, falling to below the value of 2.89 expected theoretically for a gas of 9000 K \citep{agn3,og09}. Moreover, earlier work \citep{oh65,jps73,pp77,b91} using many fewer samples showed this same trend, but the pattern had not been recognized.  The present study contains many more samples that all previous studies combined and goes out to greater distances. Our results show the same pattern, as illustrated in Figure \ref{fig:HaHb}.  Again we see the dependence of F(\Halpha)/F(\Hbeta) on distance from \ori, with values systematically falling below the theoretical values (2.89).  Following \cite{og09} and an earlier suggestion by \citet{pei82} we interpret this as meaning that at distances beyond about 5\arcmin\ scattered light from the PDR begins to be an important contributor to the total flux in these Balmer lines. The scattered light component would have to be wavelength dependent, with shorter wavelengths preferentially scattered more.  Almost certainly the explanation is not due to the flux ratio being dependent on the electron temperature since \citep{agn3} the ratio is 2.87 at 10000 K and drops to 2.69 (near the average of the furthest samples) at  the extreme temperature of 20000 K. 

\subsubsection{The Wavelength Dependence of the Scattered Light}

\begin{figure}
\epsscale{1.0} 
\plotone{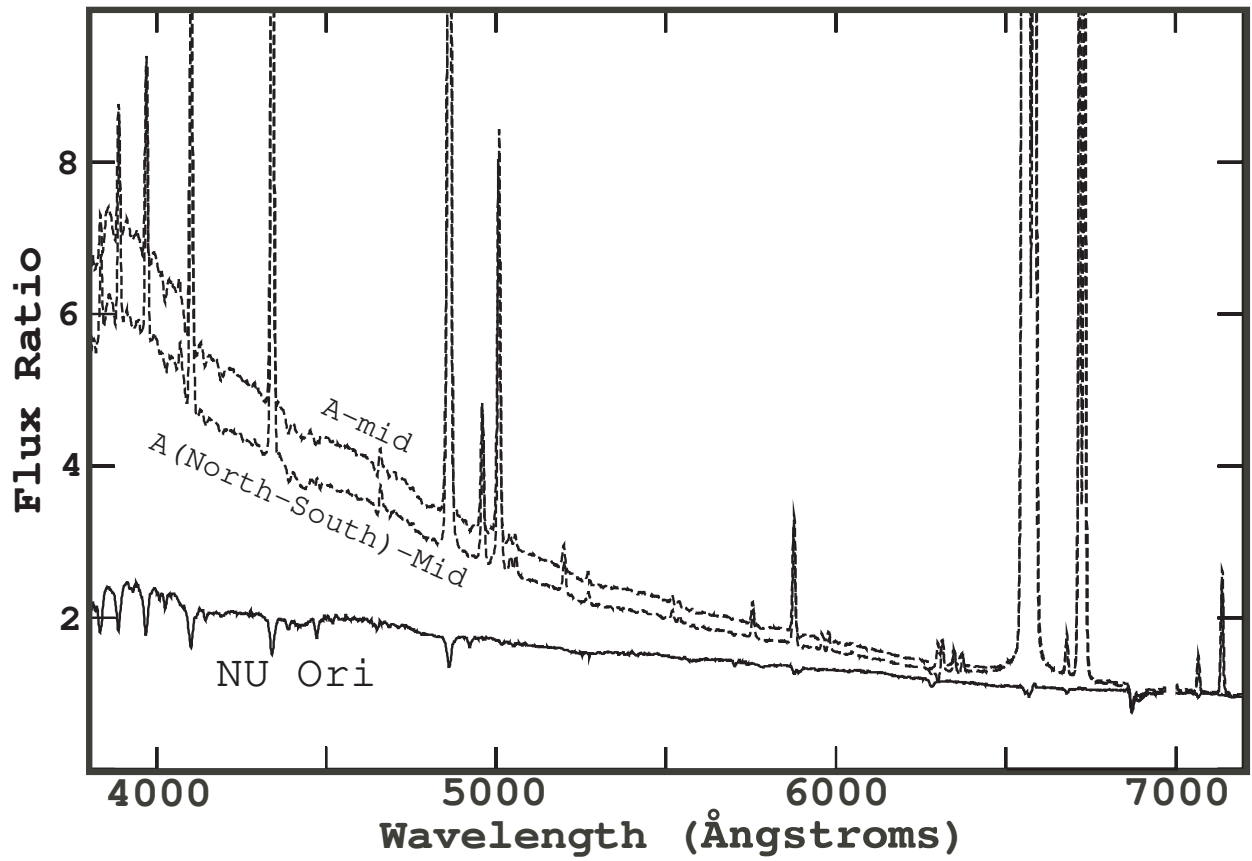}
\caption{The observed flux distribution, normalized to unity at 7000 \AA, is shown for NU Ori, and the two samples of M43 closest to the star. The interstellar reddening is very similar for each sample and this figure illustrates the strong preferential scattering of shorter wavelength radiation.
\label{fig:ThreeCont}}
\end{figure}

We can get some insight into the wavelength dependence of scattering by the PDR by comparing the scattered starlight continuum with that of the illuminating star(s).  Using the values for recombination coefficients and collision rates from \citet{agn3}we have calculated the EW for a dust-free gas of 9000 K woud be about 1600 \AA\ for \Ne =200 \cmq\ amd 1800 \AA\ for \Ne = 6000 \cmq (the difference being due to suppression of two photon emission by collisional de-excation of the 2~$\rm ^{2}$p state of hydrogen. This means that even in the innermost regions of the Huygens Region the continuum is dominated by scattered starlight. However, we don't know the spacing of any of the bright stars, with the exception of \ori, with respect to the PDR and don't know which ones are the most important contributors. We can, however, get useful information by comparing the continuum in M~43 with NU~Ori, which clearly dominates this object.  M~43 is also a blister type H~II Region with the dominant ionizing star (NU Ori) lying in front of the local ionized gas, behind which lies a PDR \citep{her97}. Once again we are dealing with a geometry involving backscattering.  In Figure \ref{fig:ThreeCont} we compare the observed flux in the A-mid and A(N-S slit)-Mid samples, which are the closest to NU~Ori, with the observed flux from NU~Ori itself. The NU~Ori observations were made during the same observing run as the M~43 observations. All of these have been normalized to be unity at 7000 \AA. The \chb\ extinction of the two samples (0.77 and 0.71) indicates a foreground reddening essentially the same as that of A$\rm _{V}$=1.74 determined by \citet{pen75} for NU Ori.  We see that the scattered light is much stronger in the blue, which strengthens the interpretation of the anomalous F(\Halpha)/F(\Hbeta) being due to a mixture of locally emitting radiation and scattered Huygens Region radiation.  There may be some reddening of the Huygens Region radiation before reaching the PDR, which means that one would expect there to be a wide scatter of  the line ratio at large distances. \citet{gh39} were the first to argue that the continuum of the Huygens Region is bluer than the the brightest stars in the region, a conclusion rendered more accurately in this study 70 years later.

\subsubsection{The Impact of Scattered Light on the Interpretation of Emission-Line Radiation}

The presence of a significant fraction of wavelength dependent scattering means that the worse fears posited in \citet{og09} and \citet{od01} seem to be confirmed.  This is true for both the bright inner region and the EON.

The Huygens Region has been studied spectroscopically at sufficiently high resolution to determine that there is backscattering of the strongest emission-lines by the PDR \citep{od92,hen94,hen98,od01} at the level of 15-20\%\ of the emitting radiation. This is revealed because there is a velocity difference between the approaching emitting layers and the stationary scattering layer, which produces a doubled velocity shift in the scattered light. The strong wavelength dependence of scattering efficiency means that one can expect that the fraction of light scattered will vary with wavelength, presumably becoming greater with decreasing wavelength. Any study that does not isolate the flux from the main emission-line and scattered light will include an important photometric error. This has always been the case, including in this study, because low velocity resolution studies and none of the high resolution studies have attempted this. Since the velocity resolution of these high resolution-broad wavelength range studies have typically been about 10 \kms\ even they may lack the necessary resolution to determine this correction. 

This effect can systematically affect the analysis of nebulae like the Orion Nebula. An example is in the determination of electron temperatures by comparing the shorter wavelength auroral lines with the nebular lines, which is commonly done for [S~II], [N~II], and [O~III]. Scattering would mean that the temperature determined from samples affected by preferential blue scattering would be too large. This effect would have an opposite effect on the reddening corrections because it would cause the extinction to be under-estimated and the auroral line26s under-corrected for reddening, leading to artificially low derived electron temperatures.  It is not clear which of these two effects would prevail. Wavelength dependent scattering could contribute to the temperature discrepancy that is known to exist when comparing the results of different methods.  A systematic over-estimate of the forbidden line temperatures would mean that the forbidden line ions actually have greater abundances. This is the sense of the abundance discrepancy that is thought to apply to H~II Regions and planetary nebulae \citep{pei09}. Scattering could, therefore, be an important contributor to the so-called t$\rm ^{2}$ effect and could potentially remove the apparent abundance discrepancy. 

The presence of large-scale scattering in the EON of Huygens Region radiation means that if one were to analyze the integrated spectrum of the entire Orion Nebula that the line ratios would be a mix of direct radiation from the Huygens Region, intrinsic radiation from the EON, and Huygens Region radiation scattered within the EON.  Such an analysis would be subject to systematic errors that cannot be removed without a thorough knowledge of the geometry of the object and the scattering properties of its particles.  At a distance of 440 pc, the 33\arcmin x 24\arcmin\ major axes of the EON corresponds to dimensions of 4.2 pc x 3.1 pc. The larger dimension would subtend 19\arcsec\ at a distance to the Large Magellanic Cloud of 46.8 kpc  \citep{wa01} and 1.20\arcsec\ at a distance of 744 kpc for the Andromeda Galaxy \citep{vi10}.  This means that characteristic determinations of abundances from emission-lines will be affected by scattering for these nearby galaxies and the effects will also be present in the study of more distance galaxies.

\section{Summary of Conclusions}

Our new spectroscopic observations have allowed us to reach multiple conclusions, some of which find application outside of the Orion Nebula.

~1. The electron temperature derived from nebular to auroral forbidden line ratios shows little change with distance from \ori, although the scattered light effect mentioned in conclusion 9 may mask a general decrease with distance. The electron density clearly decreases with increasing distance from \ori,  although again the outmost samples are probably contaminated by light scattered from the Huygens Region.

~2. The electron temperature in the inner regions, where scattering is not as important indicates that there may be a difference in the various ionization layers with [S~II] giving \Te=7390$\pm$780 K, [N~II] 8390$\pm$170 K, and [O~III] 8610$\pm$640 K which agree with the best radio data determination of 8300$\pm$200 K for the entire ionized hydrogen layer and hint that there may be a gradient in electron temperature contrary to that expected from radiation hardening. 

~3. Scattering of Huygens Region radiation by dust in the underlying PDR becomes important beyond a distance of 5\arcmin\ from \ori.  This means that the derived conditions of electron temperature and density for samples distant from \ori\ are subject to systematic error as one is actually seeing radiation from both the outer and inner regions.  This scattering preferentially favors shorter wavelengths. The wavelength dependence causes the observed F(\Halpha)/F(\Hbeta) to fall below its well-defined theoretical value at large distances from \ori\ and would cause the electron temperatures to be over-estimated.

~4. Even in the Huygens Region backscattering of emission-lines will play an important role in distorting derived properties of the nebula unless one employs a spectroscopic resolution that isolates and removes the wavelength dependent scattered light component.

~5. Diagnostic line ratios that indicate conditions of ionization, combined with measures of the extinction corrected surface brightness in \Hbeta\ and the equivalent width of the underlying continuum,  are useful for determining a 3-D model of this region. The derived model is an irregular concave surface bounded at the extremes by nearly edge-on walls lying in the foreground, but closer to \ori\ than the foreground 
Veil.  M~43 is seen to be isolated from the Huygens Region by the wall defining the outer boundary of the EON. 

~5. Many  of the optically bright H~II regions are going to be similar to the 3-D model for Orion. This means that the scattered light phenomenon will also occur there and also affect derivations of nebular conditions.

\acknowledgments
We are grateful to Jose Velasquez  of the CTIO staff for assistance in obtaining the spectra and to Robert H. Rubin  of NASA's Ames Research Center for calculating alternative values of temperatures and densities for [S~II]. Partial financial support for CRO's work on this project was provided by STScI grant GO 10967 (CRO, Principal Investigator). Jessica Harris participated in this program while a student in the Fisk University-Vanderbilt University Bridge program, supported by the National Science Foundation (PAARE grant AST-0849736).  Vanderbilt University participation in the SMARTS consortium is funded through the Vanderbilt Initiative in Data-Intensive Astrophysics.

{\it Facilities:} \facility{CTIO(1.5 m)}.
\appendix
\section{Properties of the Samples Observed}
Insert Table~\ref{tab:SampleProperties} here.

\section{Reddening Corrected Flux Ratios-One}
Insert Table~\ref{tab:FluxesFirst} here.

Insert Table~\ref{tab:FluxesSecond} here.

Insert Table~\ref{tab:FluxesThird} here.

Insert Table~\ref{tab:FluxesFourth} here.

Insert Table~\ref{tab:FluxesFifth} here.

Insert Table~\ref{tab:FluxesSixth} here.

Insert Table~\ref{tab:FluxesSeventh} here.

\section{Derived Electron Temperatures and Densities}
Insert Table~\ref{tab:TempDensities} here.

\begin{figure}
\epsscale{1.0} 
\plotone{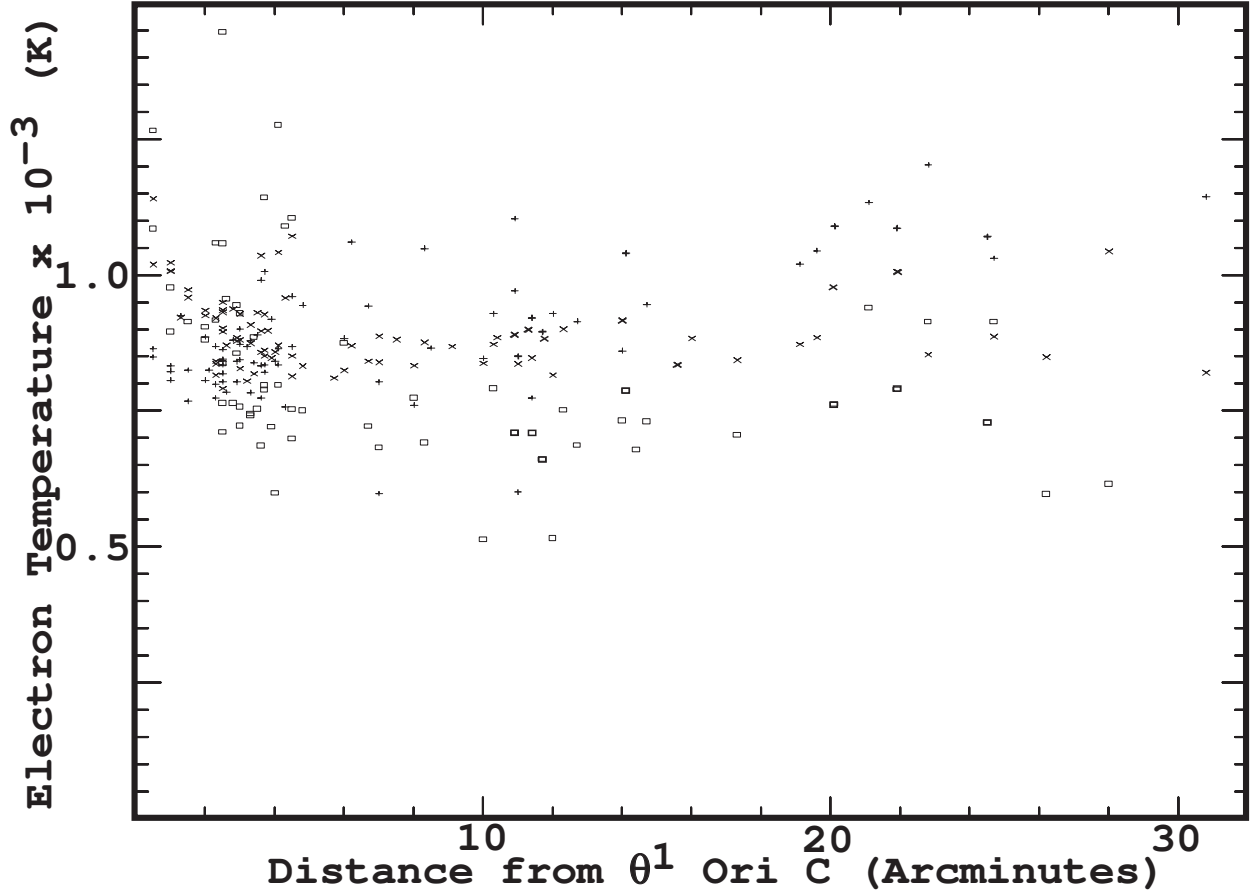}
\caption{The values of the electron temperature derived from reddening corrected auroral to nebular line ratios are shown as a function of distance from \ori. Boxes are for [S~II], x's for [N~II], and +'s for [O~III]. 
The heavy line symbols are samples from the known extended X-ray regions discussed in \S\ \ref{sect:xray}.
\label{fig:Temps}}
\end{figure}

\begin{figure}
\epsscale{1.0} 
\plotone{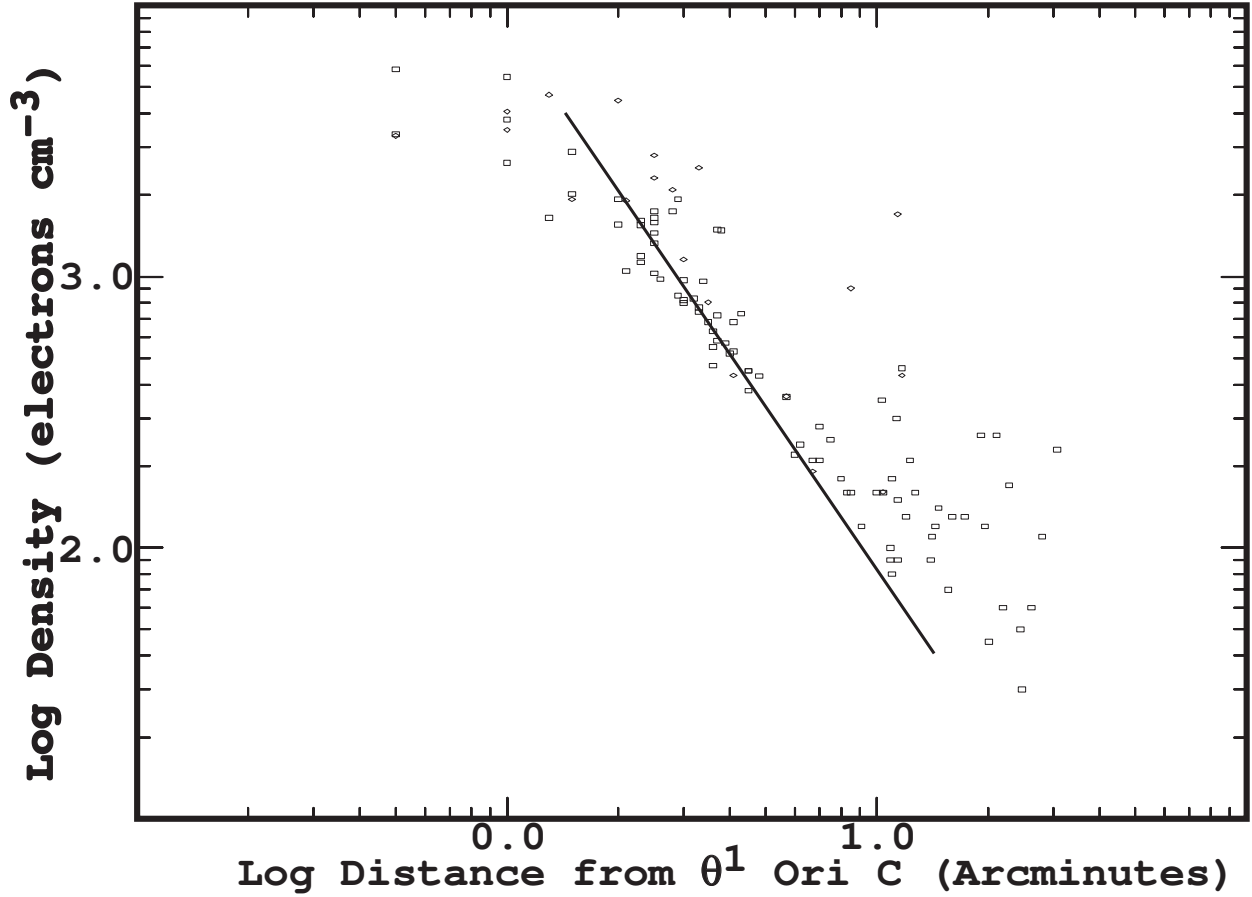}
\caption{The values of the electron density derived from the reddening corrected nebular doublet ratios are shown as a function of distance from \ori. Boxes are for [S~II] and diamonds for [Cl~III]. A line with \Ne$\sim$ D$\rm ^{-2}$ is shown for reference.
\label{fig:Densities}}
\end{figure}

\begin{deluxetable}{lcc}
\tabletypesize{\scriptsize}
\tablecaption{Comparison of Observed Line Flux Ratios with those of B1991\label{tab:B1991Comparison}}
\tablewidth{0pt}
\tablehead{
\colhead{Wavelength (\AA)}  & \colhead{B1991} & \colhead{This Study}}
\startdata
4340  &  0.45  &  0.43\\
4861  &  1.00  & 1.00\\
4959  &  0.92  & 0.93\\
5007  &  2.76  &  2.75\\
6563  & 3.34   & 3.20\\
6716  &  0.061 & 0.058\\
6731  &  0.076 & 0.068\\
7065  & 0.047  & 0.048\\
\enddata
\end{deluxetable}

\begin{deluxetable}{lcccccccc}
\tabletypesize{\scriptsize}
\tablecaption{Spectroscopic Samples
\label{tab:SampleProperties}}
\tablewidth{0pt}
\tablehead{
\colhead{Sample}      & \colhead{length} & \colhead{~~Center Position\tablenotemark{a}}&\colhead{Distance}  & \colhead{Date of}     & \colhead{Exposure}        & \colhead{\chb} & \colhead{\Shb\tablenotemark{e}} & \colhead{Equivalent}\\
\colhead{Designation} & \colhead{(arcsec)}       & \colhead{~~RA~~~~~~~DEC}            &\colhead{from Star\tablenotemark{b}} & \colhead{Observation\tablenotemark{c}} & \colhead{Time\tablenotemark{d}} & \colhead{}     & \colhead{} &\colhead{Width at  \Hbeta\ (\AA)}}

\startdata
1-east &  214 & 5:35:23.7 -5:19:23 & 4.5  &   11-25-08 &  600  &      0.51 & 5.44E-14 & 110\\
1-west &  214 &  5:35:09.3  -5:19:23 & 4.3 &    11-25-08 &  600    &    0.33 & 4.61E-14 & 170\\
2-east &  130 &  5:35:26.5  -5:19:53 & 4.5 &    11-25-08 &  120    &    0.36 & 2.93E-14 & 120\\
2-mid  &  130 &  5:35:16.5  -5:19:53 & 3.5 &    11-25-08 &  120    &    0.42 & 2.16E-13 & 270\\
2-west &  130 &  5:35:09.3  -5:19:53 & 4.1 &    11-25-08 &  120    &    0.15 & 1.33E-13 & 400\\
3-east &  130 &  5:35:26.5  -5:20:23 & 4.1 &    11-25-08 &  600    &    0.73 & 2.10E-13 & 130\\
3-mid  &  130 &  5:35:16.5  -5:20:23 & 3.0 &    11-25-08 &  600    &    0.53 & 4.13E-13 & 270\\
3-west &  130 &  5:35:09.3  -5:20:23 & 3.7 &    11-25-08 &  600    &    0.20 & 1.92E-13 & 150\\
\enddata
\tablenotetext{a}{2000}
\tablenotetext{b}{Arcmin from \ori\ except for A samples which are arcsec from NU Ori.}
\tablenotetext{c}{In month, date, year format.}
\tablenotetext{d}{Total exposure time in seconds}
\tablenotetext{e}{Extinction Corrected \Hbeta\ surface brightness in ergs \cmsq\ \ps\ \arcsecsq\ in power of ten format, e.g. 2.00E-14=2.00x10$^{-14}$}
\end{deluxetable}

\begin{deluxetable}{lccccccccccccccc}
\rotate
\setlength{\tabcolsep}{0.02in}
\tabletypesize{\scriptsize}
\tablecaption{Reddening Corrected Flux Ratios-First
\label{tab:FluxesFirst}}
\tablewidth{0pt}
\tablehead{
\colhead{Wavelength}  & \colhead{Source} & \colhead{f($\lambda)$} & \colhead{1-east} & \colhead{1-west} & \colhead{2-east} & \colhead{2-mid} & \colhead{2-west}  & \colhead{3-east} & \colhead{3-mid} & \colhead{3-west} & \colhead{4-east}&\colhead{4-mid}&\colhead{4-west}&\colhead{5-east}&\colhead{5-mid}\\ 
\colhead{(\AA)}              & \colhead{}              & \colhead{}                        & \colhead{}             & \colhead{}            & \colhead{}         & \colhead{}              & \colhead{}                 & \colhead{}             & \colhead{}           & \colhead{}             &\colhead{}             &\colhead{}          &\colhead{}            &\colhead{}           &\colhead{}
}
\startdata
3869	&	Ne[III]	&	0.145	&	0.0941	&	0.1105	&	0.1385	&	0.0903	&	0.0821	&	0.1434	&	0.0997	&	0.0889	&	0.1373	&	0.1051	&	0.0964	&	-	&	0.1065	\\
4069+76 	&	 [SII]	&	0.122	&	0.0356	&	0.0287	&	0.0267	&	0.0210	&	0.0199	&	0.0268	&	0.0278	&	0.0144	&	0.0298	&	0.0169	&	0.0126	&	-	&	0.0215	\\
4102  	&	 H I	&	0.117	&	0.2138	&	0.2295	&	0.2253	&	0.2375	&	0.2350	&	0.2244	&	0.2448	&	0.2324	&	0.2210	&	0.2496	&	0.2290	&	0.2481	&	0.2421	\\
4340	&	 Hgamma	&	0.086	&	0.4607	&	0.4614	&	0.4755	&	0.4731	&	0.4585	&	0.4641	&	0.4744	&	0.4626	&	0.4594	&	0.4723	&	0.4623	&	0.4660	&	0.4746	\\
4363	&	 [O III]	&	0.078	&	0.0158	&	0.0047	&	0.0114	&	0.0066	&	0.0053	&	0.0080	&	0.0069	&	0.0058	&	0.0151	&	0.0076	&	0.0079	&	-	&	0.0090	\\
4471	&	 He I	         &	0.064	&	0.0068	&	0.0349	&	0.0402	&	0.0360	&	0.0388	&	0.0330	&	0.0389	&	0.0405	&	0.0359	&	0.0414	&	0.0402	&	-	&	0.0431	\\
4658	&	[FeII]      	&	0.033	&	0.0075	&	0.0075	&	0.0104	&       0.0104	&	0.0102	&	0.0063	&	0.0116	&	0.0103	&	0.0076	&	0.0113	&	0.0094	&	-	&	0.0134	\\
4861	&	Hbeta	&	0.000	&	1.0000	&	1.0000	&	1.0000	&	1.0000	&	1.0000	&	1.0000	&	1.0000	&	1.0000	&	1.0000	&	1.0000	&	1.0000	&	1.0000	&	1.0000	\\
\enddata
\tablenotetext{a}{Small contributions from [Fe~II] at 5262 \AA\ and O~I at 5275 \AA. The 11-25-08 observations had a bad pixel column that prevented measuring these lines.}
\end{deluxetable}

\begin{deluxetable}{lccccccccccccccc}
\rotate
\setlength{\tabcolsep}{0.02in}
\tabletypesize{\scriptsize}
\tablecaption{Reddening Corrected Flux Ratios-Second
\label{tab:FluxesSecond}}
\tablewidth{0pt}
\tablehead{
\colhead{Wavelength}  & \colhead{5-west} & \colhead{6-east} & \colhead{6-mid} & \colhead{6-west} & \colhead{7-east} & \colhead{7-mid} & \colhead{7-west}  & \colhead{8-east} & \colhead{8-mid} & \colhead{8-west} & \colhead{9-east}&\colhead{9-mid}&\colhead{9-west}&\colhead{10-east}&\colhead{10-mid}\\ 
\colhead{(\AA)}              & \colhead{}              & \colhead{}                        & \colhead{}             & \colhead{}            & \colhead{}         & \colhead{}              & \colhead{}                 & \colhead{}             & \colhead{}           & \colhead{}             &\colhead{}             &\colhead{}          &\colhead{}            &\colhead{}           &\colhead{}
}
\startdata
3869	&	0.0911	&	-	&	0.1230	&	0.1062	&	0.1185	&	0.1550	&	0.0988	&	0.1429	&	0.1887	&	0.1219	&	-	&	-	&	-	&	0.1038	&	0.1861	\\
4069+76 	&	0.0033	&	-	&	0.0053	&	0.0169	&	0.0271	&	0.0192	&	0.0138	&	-	&	0.0177	&	0.0193	&	-	&	-	&	-	&	0.0154	&	0.0310	\\
4102	&	0.2365	&	-	&	0.2509	&	0.2418	&	0.2604	&	0.2542	&	0.2555	&	0.2637	&	0.2505	&	0.2483	&	-	&	-	&	-	&	0.2371	&	0.2600	\\
4340	&	0.4694	&	0.4949	&	0.4924	&	0.4724	&	0.5279	&	0.4928	&	0.4791	&	0.4798	&	0.4905	&	0.4647	&	0.4852	&	0.4698	&	0.4754	&	0.4628	&	0.4706	\\
4363	&	0.0075	&	0.0121	&	0.0079	&	0.0068	&	0.0080	&	0.0093	&	0.0072	&	-	&	0.0129	&	0.0102	&	0.0157	&	0.0103	&	0.0058	&	0.0060	&	0.0100	\\
4471	&	0.0377	&	0.0607	&	0.0449	&	0.0393	&	0.0403	&	0.0440	&	0.0508	&	0.0410	&	0.0476	&	0.0416	&	0.0462	&	0.0444	&	0.0429	&	0.0412	&	0.0454	\\
4658	&	0.0088	&	-         	&	0.0124	&	0.0095	&	0.0100	&	0.0136	&	0.0093	&	0.0133	&	0.0109	&	0.0110	&	0.0077	&	0.0091	&	0.0076	&	0.0094	&	0.0095	\\
4861	&	1.0000	&	1.0000	&	1.0000	&	1.0000	&	1.0000	&	1.0000	&	1.0000	&	1.0000	&	1.0000	&	1.0000	&	1.0000	&	1.0000	&	1.0000	&	1.0000	&	1.0000	\\
\enddata
\end{deluxetable}

\begin{deluxetable}{lccccccccccccccc}
\rotate
\setlength{\tabcolsep}{0.02in}
\tabletypesize{\scriptsize}
\tablecaption{Reddening Corrected Flux Ratios-Third
\label{tab:FluxesThird}}
\tablewidth{0pt}
\tablehead{
\colhead{Wavelength}  & \colhead{10-west} & \colhead{11-east} & \colhead{11-mid} & \colhead{11-west} & \colhead{12-east} & \colhead{12-mid} & \colhead{12-west)}  & \colhead{13-east} & \colhead{13-mid} & \colhead{13-west} & \colhead{14-east}&\colhead{14-mid}&\colhead{14-west}&\colhead{15-east}&\colhead{15-mid}\\ 
\colhead{(\AA)}              & \colhead{}              & \colhead{}                        & \colhead{}             & \colhead{}            & \colhead{}         & \colhead{}              & \colhead{}                 & \colhead{}             & \colhead{}           & \colhead{}             &\colhead{}             &\colhead{}          &\colhead{}            &\colhead{}           &\colhead{}
}
\startdata
3869	&	0.1417	&	0.0934	&	0.1700	&	0.1378	&	0.0647	&	0.0139	&	0.1310	&	0.0948	&	0.0920	&	0.0986	&	0.0626	&	0.0824	&	0.0873	&	0.0602	&	0.0789	\\
4069+76 	&	0.0155	&	0.0218	&	0.0178	&	0.0133	&	0.0177	&	0.0186	&	0.0099	&	0.0235	&	0.0151	&	0.0169	&	0.0134	&	0.0215	&	0.0126	&	0.0143	&	0.0161	\\
4102	&	0.2500	&	0.2441	&	0.2503	&	0.2535	&	0.2319	&	0.2548	&	0.2549	&	0.2212	&	0.2503	&	0.2533	&	0.2327	&	0.2447	&	0.2444	&	0.2335	&	0.2476	\\
4340	&	0.4762	&	0.4843	&	0.4751	&	0.4840	&	0.4643	&	0.4911	&	0.4818	&	0.4652	&	0.4731	&	0.4852	&	0.4654	&	0.4827	&	0.4755	&	0.4629	&	0.4836	\\
4363	&	0.0076	&	0.0071	&	0.0088	&	0.0067	&	0.0048	&	0.0065	&	0.0071	&	0.0038	&	0.0088	&	0.0070	&	0.0039	&	0.0055	&	0.0052	&	0.0042	&	0.0049	\\
4471	&	0.0442	&	0.0400	&	0.0439	&	0.0424	&	0.0252	&	0.0450	&	0.0430	&	0.0267	&	0.0457	&	0.0450	&	0.0335	&	0.0378	&	0.0427	&	0.0356	&	0.0375	\\
4658	&	0.0076	&	0.0124	&	0.0092	&	0.0058	&	0.0091	&	0.0093	&	0.0086	&	0.0126	&	0.0081	&	0.0084	&	0.0107	&	0.0101	&	0.0086	&	0.0092	&	0.0076	\\
4861	&	1.0000	&	1.0000	&	1.0000	&	1.0000	&	1.0000	&	1.0000	&	1.0000	&	1.0000	&	1.0000	&	1.0000	&	1.0000	&	1.0000	&	1.0000	&	1.0000	&	1.0000	\\
\enddata
\end{deluxetable}

\begin{deluxetable}{lccccccccccccccc}
\rotate
\setlength{\tabcolsep}{0.02in}
\tabletypesize{\scriptsize}
\tablecaption{Reddening Corrected Flux Ratios-Fourth
\label{tab:FluxesFourth}}
\tablewidth{0pt}
\tablehead{
\colhead{Wavelength}  & \colhead{15-west} & \colhead{16-east} & \colhead{16-mid} & \colhead{16-west} & \colhead{17}          & \colhead{18}          & \colhead{19-east}  & \colhead{19-west} & \colhead{20}          & \colhead{21}          & \colhead{22} & \colhead{23} & \colhead{24-NE} & \colhead{24-SW} & \colhead{25-NE}\\ 
\colhead{(\AA)}               & \colhead{}               & \colhead{}               & \colhead{}             & \colhead{}                & \colhead{}               & \colhead{}              & \colhead{}                & \colhead{}               & \colhead{}               & \colhead{ }              & \colhead{}      & \colhead{}      & \colhead{}            & \colhead{}              & \colhead{}
}
\startdata
3869	&	0.0750	&	0.0439	&	0.0706	&	0.0934	&	0.0855	&	0.0632	&	-         	&	-	        &	0.0518	&	0.0493	&	0.0514	&	0.0474	&	-	         &	-	         &	-	       \\
4069+76&0.0180	&	0.0184	&	0.0125	&	0.0185	&	0.0188	&	0.0124	&	-	         &	-	        &	0.0163	&	0.0097	&	0.0111	&	0.0234	&	-	         &	-	         &	-	        \\
4102	&	0.2385	&	0.2325	&	0.2441	&	0.2377	&	0.2306	&	0.2166	&       	-	&	-	         &	0.2199	&	0.2131	&	0.2189	&	0.2105	&	-	         &	-	         &	-	        \\
4340	&	0.4631	&	0.4624	&	0.4813	&	0.4742	&	0.4714	&	0.4480	&	0.4567	&	0.4707	&	0.4507	&	0.4477	&	0.4529	&	0.4456	&	0.4649	&	0.4643	&	0.4649     \\
4363	&	0.0054	&	0.0038	&	0.0036	&	0.0061	&	0.0049	&	0.0035	&	-	         &	-	         &	0.0023	&	0.0027	&	0.0044	&	0.0045	&	0.0086	&	0.0039	&	0.0114	\\
4471	&	0.0367	&	0.0224	&	0.0278	&	0.0368	&	0.0295	&	0.0286	&	0.0258	&	0.0136	&	0.0264	&	0.0195	&	0.0163	&	0.0180	&	0.0417	&	0.0344	&	0.0309	\\
4658	&	0.0120	&	0.0097	&	0.0071	&	0.0070	&	0.0067	&	0.0093	&	0.0066	&	0.0070	&	0.0062	&	0.0064	&	0.0072	&	0.0043	&	0.0071	&	0.0076	&	0.005	\\
4861	&	1.0000	&	1.0000	&	1.0000	&	1.0000	&	1.0000	&	1.0000	&	1.0000	&	1.0000	&	1.0000	&	1.0000	&	1.0000	&	1.0000	&	1.0000	&	1.0000	&	1.0000	\\
\enddata
\end{deluxetable}

\begin{deluxetable}{lccccccccccccccc}
\rotate
\setlength{\tabcolsep}{0.02in}
\tabletypesize{\scriptsize}
\tablecaption{Reddening Corrected Flux Ratios-Fifth
\label{tab:FluxesFifth}}
\tablewidth{0pt}
\tablehead{
\colhead{Wavelength}  & \colhead{25-SW} & \colhead{26-n} & \colhead{26-s} & \colhead{27-east} & \colhead{27-west} & \colhead{28-east} & \colhead{28-west}  & \colhead{A-east} & \colhead{A-mid} & \colhead{A-rift} & \colhead{A-west}&\colhead{A-north}&\colhead{A(N-S)-mid}&\colhead{A-south}&\colhead{B}\\ 
\colhead{(\AA)}              & \colhead{}              & \colhead{}                        & \colhead{}             & \colhead{}            & \colhead{}         & \colhead{}              & \colhead{}                 & \colhead{}             & \colhead{}           & \colhead{}             &\colhead{}             &\colhead{}          &\colhead{}            &\colhead{}           &\colhead{}
}
\startdata
3869	&	-	         &	-	        &	-	         &	-	         &	-	         &	-	         &	-	         &	0.1038	&	0.0612	&	-	         &	-	         &	-	         &	-	         &	-	          &	0.0554	\\
4069+76 &	-	&	-	        &	-	         &	-	         &	-	         &	-	         &	-	         &	0.0368	&	-	          &	-	         &	0.0305	&	0.0457	&	0.0279	&	0.0463	&	0.0162	\\
4102	&	-          	&	-	        &	-	         &	-	         &	-	         &	-	         &	-	         &	0.2377	&	0.1912	&	0.1842	&	0.2149	&	0.2174	&	0.2029	&	0.2071	&	0.2324	\\
4340	&	0.4650	&	0.4650     &	0.4580	&	0.4591	&	0.4508	&	0.4589	&	0.4395	&	0.4632	&	0.4329	&	0.4198	&	0.4404	&	0.4444	&	0.4357	&	0.4333	&	0.4467	\\
4363	&	0.0022	&	-	        &	-	         &	0.0040	&	0.0051	&	-	          &	0.0118	&	-	         &	-	        &	-	          &	-	         &	-	          &	-	         &	-	         &	0.0051	\\
4471	&	0.0270	&	-	        &	0.0333	&	0.0391	&	0.0305	&	0.0245	&	0.0329	&	0.0133	&	-	         &	-	         &	-	         &	-	          &	-	          &	-	         &	0.0368	\\
4658	&	0.0097	&	-	        &	0.0052	&	0.0078	&	0.0049	&	0.0059	&      0.0045	&	0.0084	&	0.0066	&	0.0097	&	0.0078	&	0.0061	&	0.0080	&	0.0062	&	0.0765	\\
4861	&	1.0000	&	1.0000	&	1.0000	&	1.0000	&	1.0000	&	1.0000	&	1.0000	&	1.0000	&	1.0000	&	1.0000	&	1.0000	&	1.0000	&	1.0000	&	1.0000	&	1.0000	\\
\enddata
\end{deluxetable}

\begin{deluxetable}{lccccccccccccccc}
\rotate
\setlength{\tabcolsep}{0.02in}
\tabletypesize{\scriptsize}
\tablecaption{Reddening Corrected Flux Ratios-Sixth
\label{tab:FluxesSixth}}
\tablewidth{0pt}
\tablehead{
\colhead{Wavelength}  & \colhead{B1991} & \colhead{C} & \colhead{D-NE} & \colhead{D-NW} & \colhead{D-SE} & \colhead{D-SW} & \colhead{D-north}  & \colhead{D-south} & \colhead{E} & \colhead{F-north} & \colhead{F-south}&\colhead{G}&\colhead{H}&\colhead{I}&\colhead{J}\\ 
\colhead{(\AA)}              & \colhead{}              & \colhead{}                        & \colhead{}             & \colhead{}            & \colhead{}         & \colhead{}              & \colhead{}                 & \colhead{}             & \colhead{}           & \colhead{}             &\colhead{}             &\colhead{}          &\colhead{}            &\colhead{}           &\colhead{}
}
\startdata
3869	&	0.1083	&	0.0521	&	0.0440	&	0.0318	&	0.0483	&	0.0376	&	0.0303	&	0.0367	&	0.0594	&	0.0397	&	0.0433	&	0.0711	&	0.0693	&	0.0867	&	0.1020	\\
4069+76 	&	0.0125	&	0.0209	&	0.0166	&	0.0317	&	0.0190	&	0.0300	&	0.0253	&	0.0302	&	0.0335	&	0.0550	&	0.0278	&		&	0.0415	&	0.0404	&	0.0212	\\
4102	&	0.2364	&	0.2291	&	0.2351	&	0.2392	&	0.2335	&	0.2356	&	0.2390	&	0.2283	&	0.2427	&	0.2347	&	0.2173	&	0.2040	&	0.2387	&	0.2142	&	0.1998	\\
4340	&	0.4610	&	0.4528	&	0.4527	&	0.4534	&	0.4501	&	0.4442	&	0.4546	&	0.4497	&	0.4696	&	0.4585	&	0.4236	&	0.4399	&	0.4492	&	0.4571	&	0.4234	\\
4363	&	0.0088	&	0.0041	&	0.0057	&	0.0027	&	0.0047	&	0.0037	&	0.0035	&	0.0053	&	0.0092	&	0.0040	&		&	0.0128	&	0.0090	&	0.0080	&		\\
4471	&	0.0442	&	0.0073	&	0.0376	&	0.0299	&	0.0357	&	0.0283	&	0.0375	&	0.0211	&	0.0205	&	0.0162	&	0.0197	&	0.0274	&	0.0159	&	0.0177	&	0.0212	\\
4658	&	0.0076	&	0.0073	&	0.0065	&	0.0049	&	0.0054	&	0.0041	&	0.0045	&	0.0071	&	0.0058	&	0.0058	&	0.0053	&	0.0074	&	0.0049	&	0.0036	&	0.0078	\\
4861	&	1.0000	&	1.0000	&	1.0000	&	1.0000	&	1.0000	&	1.0000	&	1.0000	&	1.0000	&	1.0000	&	1.0000	&	1.0000	&	1.0000	&	1.0000	&	1.0000	&	1.0000	\\
\enddata
\end{deluxetable}

\begin{deluxetable}{lccccccccccc}
\rotate
\setlength{\tabcolsep}{0.02in}
\tabletypesize{\scriptsize}
\tablecaption{Reddening Corrected Flux Ratios-Seventh
\label{tab:FluxesSeventh}}
\tablewidth{0pt}
\tablehead{
\colhead{Wavelength}  & \colhead{JW831} & \colhead{JW873-SW} & \colhead{K-east} & \colhead{K-west} & \colhead{L-east} & \colhead{L-west} & \colhead{M4}  & \colhead{V1-east} & \colhead{V1-west} & \colhead{V3-east} & \colhead{V3-west}\\ 
\colhead{(\AA)}              & \colhead{}              & \colhead{}                        & \colhead{}             & \colhead{}            & \colhead{}         & \colhead{}              & \colhead{}                 & \colhead{}             & \colhead{}           & \colhead{}             &\colhead{}            
}
\startdata
3869	&	-	&	-	&	0.0668	&	0.0551	&	0.0756	&	0.0731	&	0.0326	&	0.0804	&	0.0316	&	0.0666	&	0.0327	\\
4069+76 	&	-	&	-	&	0.0288	&	0.0286	&	0.0000	&	0.0231	&	0.0215	&	0.0215	&	0.0236	&	0.0197	&	0.0332	\\
4102	&	-	&	-	&	0.1987	&	0.2080	&	0.2096	&	0.2114	&	0.2316	&	0.2092	&	0.2352	&	0.2145	&	0.2324	\\
4340	&	0.4660	&	0.4560	&	0.4425	&	0.4332	&	0.4337	&	0.4403	&	0.4497	&	0.4304	&	0.4480	&	0.4446	&	0.4515	\\
4363	&	-	&	-	&	0.0176	&	0.0156	&	0.0120	&	-	&	0.0033	&	0.0085	&	0.0045	&	0.0088	&	0.0064	\\
4471	&	0.0284	&	0.0091	&	0.0234	&	0.0244	&	0.0217	&	0.0199	&	0.0185	&	0.0255	&	0.0135	&	0.0245	&	0.0115	\\
4658	&	0.0083	&	0.0073	&	0.0091	&	-	&	0.0067	&	0.0060	&	0.0080	&	0.5760	&	0.0083	&	0.0071	&	0.0075	\\
4861	&	1.0000	&	1.0000	&	1.0000	&	1.0000	&	1.0000	&	1.0000	&	1.0000	&	1.0000	&	1.0000	&	1.0000	&	1.0000	\\
\enddata
\end{deluxetable}

\begin{deluxetable}{lccccc}
\tabletypesize{\scriptsize}
\tablecaption{Derived Electron Temperatures and Densities\tablenotemark{a}
\label{tab:TempDensities}}
\tablewidth{0pt}
\tablehead{
\colhead{Sample}\\
\colhead{Designation} & \colhead{\Te([S~II])}       & \colhead{\Te([N~II])}            &\colhead{\Te([O~OIII])} & \colhead{\Ne([S~II])} & \colhead{\Ne([Cl~III])}}
\startdata
1-East	&	7530	&	8120	&		&	380	&		\\
1-West	&	10900	&	9560	&	7580	&	730	&		\\
2-East	&	11050	&	10700	&	9620	&	450	&		\\
2-Mid	&	7540	&	9290	&	8910	&	680	&	800	\\
2-West	&	7980	&	8690	&	8360	&	530	&		\\
3-East	&	12760	&	10400	&	8660	&	680	&	430	\\
3-Mid	&	9300	&	9270	&	8740	&	970	&	1150	\\
3-West	&	7890	&	8600	&	8220	&	580	&		\\
\enddata
\tablenotetext{a}{All temperatures are in K and densities in \cmq.}
\end{deluxetable}
\clearpage 


\clearpage
\begin{thebibliography}{}
\bibitem[Abel et~al.(2006)]{abel} Abel, N. P., Ferland, G. J., O'Dell, C. R., Shaw, G., \& Troland, T. H. \apj, 644, 344
\bibitem[Baldwin et~al.(1991)]{b91} Baldwin, J. A., Ferland, G. J., Martin, P. G., Corbin, M. R., Cota, S. A., Peterson, B. M., \& Slettebak, A. 1991, \apj, 374, 580
\bibitem[Baldwin et~al.(2000)]{b00} Baldwin, J. A., Verner, E. M., Verner, D. A., Ferland, G. J., Martin, P. G., Korista, K. T., \& Rubin, R. H. 2000, \apjs, 129, 229
\bibitem[Blagrave et~al.(2007)]{bla} Blagrave, K. P. M., Martin, P. G., Rubin, R. H., Dufour, R. J., Baldwin, J. A., Hester, J. J., \& Walter, D. K. 2007, \apj, 655, 299
\bibitem[Balick et~al.(1974)]{bal74} Balick, B., Gammon, R. H., \& Hjellming, R. M. 1974, PASP, 86, 616
\bibitem[Bally et~al.(2001a)]{jb01a} Bally, J., Johnstone, D., Joncas, G., Reipurth, B., \& Mall\'en-Omelas, G. 2001, \aj, 122, 1508
\bibitem[Bally et~al.(2001b)]{jb01b} Bally, J., O'Dell, C. R., \& McCaughrean, M. J. 2001, \aj, 119, 2919
\bibitem[Caplan (1972)]{ca72} Caplan, J. G. 1972, \aap, 18, 408
\bibitem[Deharveng (1973)]{deh73} Deharveng, L. 1973, \aap, 29, 341
\bibitem[Dicker et~al.(2009)]{dicker} Dicker, S. R., Mason, B. S., Korngut, P. M., Cotton, W. D., Compi\'egne, M., Devlin, M. J., Ade, P. A. R., Benford, D. J., Irwin, K. D., Madalena, R. J., McMullin, J. P., Shepherd, D. S., Sievers, A., Staguhn, J. G., \& Tucker, C. 2009, \apj, 705, 226
\bibitem[Dufour \& Mathis(1975)]{rjd} Dufour, R. J., \& Mathis, J. S. 1975, \pasp, 87, 345
\bibitem[Esteban et~al.(2004)]{est04} Esteban, C., Peimbert, M., Garc{\'{\i}}a-Rojas, J., Ruiz, M. T., Peimbert, A., \& Rodr{\'{\i}}guez, M. 2004, \mnras, 355, 229
\bibitem[Ferland et~al.(1998)]{fer98} Ferland, G. J., Korista, K. T., Verner, D. A., Ferguson, J. W., Kingdon, J. B., \& Verner, E. M. 1991, PASP, 110, 76
\bibitem[Garc{\'{\i}}a-D{\'{\i}}az  \& Henney(2007)]{gar07} Garc{\'{\i}}a-D{\'{\i}}az, Ma. T.,  and Henney, W. J. 2007, \aj, 133, 952
\bibitem[Garc{\'{\i}}a-D{\'{\i}}az et~al.(2008)]{gar08} Garc{\'{\i}}a-D{\'{\i}}az, Ma. T., Henney, W. J., L\'opez, J. A., \& Doi, T. 2008, RMxAA, 44, 181
\bibitem[Gingerich(1982)]{gin} Gingerich, O.  1982 Ann. NY Acad.Sci., 395, 308
\bibitem[G\"udel et~al.(2008)]{gud} G\"udel, M., Briggs, K. R., Montmerle, T., Audard, M., Rebull, L., \& Skinner, S. L. 2008, Science, 319, 309
\bibitem[Habing \& Israel(1979)]{habing} Habing, H. J., \& Israel, F. P. 1979, ARAA, 17, 345
\bibitem[Hamidouche et~al.(2008)]{ham} Hamidouche, M., Wang, S., \& Looney, L. W. 2008, AJ, 135, 1474
\bibitem[H\"anel (1987)]{han87} H\"anel, A. 1987, \aap, 176, 347
\bibitem[Henney (1994)]{hen94} Henney, W. J. 1994, \apj, 427, 288
\bibitem[Henney (1998)]{hen98} Henney, W. J. 1998, \apj, 503, 760
\bibitem[Henney et~al. (2005)]{hen05} Henney, W. J., Arthur, S. J., \& Garc{\'{\i}}a-D{\'{\i}}az, Ma. T. 2005, \apj, 627, 813\bibitem[Henney et~al.(2007)]{wjh} Henney, W. J., O'Dell, C. R., Zapata, L. A., Garc{\'{\i}}a-D{\'{\i}}az, Ma. T., Rodr{\'{\i}}guez, L. F., \& Robberto, M. 2007, \aj, 133, 2192
\bibitem[Greenstein \& Henyey(1939)]{gh39} Greenstein, J. L., \& Henyey, L. G. 1939, \apj, 89, 653
\bibitem[Herrmann et~al.(1997)]{her97} Herrmann, F., Madden, S. C., Nikola, T., Poglitsch, A.,  Timmermann, R., Geis, N., Townes, C. H., \& Stacyl. G. J. 1997, \apj, 481, 343
\bibitem[Johnson (1965)]{hmj} Johnson, H. M. 1965, ApJ, 142, 964
\bibitem[Jones \& Walker(1988)]{jw88} Jones, B. F., \& Walker, M. F. 1985, \aj, 95, 1755
\bibitem[Kaler (1967)]{kaler67} Kaler, J. B. 1967, \apj, 148, 925
\bibitem[Khallessee et~al.(1980)]{kha80} Khallesse, B., Pallister, W. S., Warren-Smith, R. F., \& Scarrott, S. M. 1980, MNRAS, 190, 99
\bibitem[Massey \& Foltz(2000)]{mf00} Massey, P., \& Folz, C. B. 2000, \pasp, 112, 556
\bibitem[O'Dell (1998)]{od98} O'Dell, C. R. 1998, \aj, 116, 1346
\bibitem[O'Dell (1999)]{od99} O'Dell, C. R. 1999, \apj, 525, 321
\bibitem[O'Dell (2001)]{od01} O'Dell, C. R. 2001, ARAA, 39, 99
\bibitem[O'Dell(2004)]{od04} O'Dell, C. R. 2004, \pasp, 116, 729
\bibitem[O'Dell(2009)]{od09} O'Dell, C. R. 2009, \pasp, 121, 428
\bibitem[O'Dell \& Goss(2009)]{og09} O'Dell, C. R. \& Goss, W. M. 2009, \aj, 138, 1235
\bibitem[O'Dell \& Henney(2008)]{oh08} O'Dell, C. R., \& Henney, W. J. 2008, \aj, 136, 1566
\bibitem[O'Dell et~al.(2009a)]{od09a} O'Dell, C. R., Henney, W. J., Abel, N. P., Ferland, G. J., \& Arthur, S. J. 2009, \aj, 137, 367
\bibitem[O'Dell et~al.(2009b)]{od09b} O'Dell, C. R., Henney, W. J., \& Sabbadin, F. 2009, \aj, 137, 3815
\bibitem[O'Dell \& Hubbard(1965)]{oh65} O'Dell, C. R., \& Hubbard, W. B. 1965, \apj, 142, 591
\bibitem[O'Dell et~al.(1992)]{od92} O'Dell, C. R., Walter, D. K., \&  Dufour, R. J. 1992, \aj, 399, L67
\bibitem[O'Dell \& Wen(1994)]{ow94} O'Dell, C. R., \& Wen, Z. 1994, \apj, 436, 194
\bibitem[O'Dell \& Wong(1996)]{ow96} O'Dell, C. R., \& Wong, S.-K. 1996, \aj, 111,846
\bibitem[O'Dell \& Yusef-Zadeh(2000)]{oyz} O'Dell, C. R., \& Yusef-Zadeh, F.  2000, \aj, 120, 382
\bibitem[Osterbrock \& Ferland(2006)]{agn3} Osterbrock, D. E., \& Ferland, G. J. 2006, Astrophysics of Gaseous Nebulae and Active Galactic Nuclei (University Science Books, Mill Valley,  CA)
\bibitem[Osterbrock et~al.(1992)]{deo92} Osterbrock, D. E., Tran, H. D., \& Veilleux, S. 1992, \apj, 389, 305 
\bibitem[Peimbert (1967)]{pei67} Peimbert, M. 1967, \apj, 150, 825
\bibitem[Peimbert (1982)]{pei82} Peimbert, M. 1982, Ann. NY Acad.Sci., 395, 24
\bibitem[Peimbert \& Costero(1969)]{peicos69} Peimbert, M., \& Costero, R. 1969, Bul.Obs.Tonantzintla, 5, 3
\bibitem[Peimbert \& Goldsmith(1972)]{man72} Peimbert, M., \& Goldsmith, D. W. 1972, \aap, 19, 398
\bibitem[Peimbert \& Peimbert(2009)]{pei09} Peimbert, M., \& Peimbert, A. 2009, arXiv:0912.23781v1
\bibitem[Peimbert \&Torres-Peimbert(1977)]{pp77} Peimbert, M., \& Torres-Peimbert, S. 1977, MNRAS, 179, 217
\bibitem[Penston et~al(1975)]{pen75} Penston, M V, Hunter, J. K., \& O'Neill, A. 1975, \mnras, 171, 219
\bibitem[Rodr{\'{\i}}guez (1999)]{rod99} Rodr{\'{\i}}guez, M. 1999, \aap, 348, 222
\bibitem[Rubin et~al.(2003)]{rubin03} Rubin. R. H., Martin, P. G., Dufour, R. J., Ferland, G. J., Blagrave, K. P. M., Liu, X.-W., Nguyen, J. F., \& Baldwin, J. A. 2003, \mnras, 340, 362
\bibitem[Schild \& Chaffee(1971)]{sch71} Schild, R. E., \& Chaffee, F. 1971, \apj, 169, 529
\bibitem[Smith et~al.(1987)]{sm87} Smith, J., Harper, D. A., \& Loewenstein, R. F. 1987, \apj, 314, 76
\bibitem[Shaw \& Dufour(1994)]{sd94} Shaw, R. A., \& Dufour, R. J. 1994, ASP Conf. Series, 61, 327
\bibitem[Simpson (1973)]{jps73} Simpson, J. P. 1977, PASP, 85, 479
\bibitem[Str\"omgren (1939)]{strom} Str\"omgren, B. 1939, \apj, 89, 526
\bibitem[Tenorio-Tagle (1979)]{tentag} Tenorio-Tagle, G.  1979, \aap, 7, 59
\bibitem[Thum et~al.(1978)]{thu78} Thum, C., Lemke, D., Fahrbach, U., \& Frey, A. 1978, \aap, 65, 207
\bibitem[van der Werf \& Goss(1989)]{vdw89} van der Werf, P., \& Goss, W. M. 1989, \aap , 224, 209
\bibitem[Vilardell et~al.(2010)]{vi10} Vilardell, F., Ribas, I., Jordi, C., Fitzpatrick, E. L., \& Guinan, E. F. 2010, \aap, 509, 70
\bibitem[Walker et~al.(2001)]{wa01}Walker, A. R., Raimondo, G., Di Carlo, E., Brocato, E., Castellani, V. \& Hill, B. 2001, \apj, 560, L139
\bibitem[Warren \& Hesser(1977)]{wh77} Warren, W. H. Jr., \& Hesser, J. E. 1977, \apjs,  34, 115
\bibitem[Wen \& O'Dell(1995)]{wo95} Wen, Z., \& O'Dell, C. R. 1995, \apj, 438, 784.
\bibitem[Wilson et~al.(1997)]{wil97} Wilson, T. L., Filges, L., Codella, C., Reich, W., \& Reich, P. 1997, \aap, 327, 1177
\bibitem[Zuckerman (1973)]{z73} Zuckerman, B. 1973, \apj, 183, 863
\end{thebibliography}
\end{document}